# Evolution of 21st Century Sea-level Rise Projections

Andra J. Garner[a,b,e*], Jeremy L. Weiss[c], Adam Parris[d], Robert E. Kopp[b,e], Radley M. Horton[f], Jonathan T. Overpeck[g], and Benjamin P. Horton[a,b,h,i]

*Corresponding Author:*
Andra J. Garner
Department of Earth and Planetary Sciences
Rutgers University
610 Taylor Road
Piscataway, NJ 08854
848-932-3482
ajgarner@marine.rutgers.edu

[a]*Department of Marine and Coastal Sciences, Rutgers University, New Brunswick, NJ 08901*
[b]*Institute of Earth, Ocean, and Atmospheric Sciences, Rutgers University, New Brunswick, NJ 08901*
[c]*School of Natural Resources and the Environment, University of Arizona, Tucson, AZ 85721*
[d]*The Science and Resilience Institute, Brooklyn College, City University of New York, Brooklyn, NY 11210*
[e]*Department of Earth and Planetary Sciences, Rutgers University, Piscataway, NJ 08854*
[f]*Lamont-Doherty Earth Observatory, Columbia University Earth Institute, Columbia University, Palisades, NY 10964*
[g]*School for Environment and Sustainability, University of Michigan, Ann Arbor, MI 48109*
[h]*Asian School of the Environment, Nanyang Technological University, Singapore 639798*
[i]*Earth Observatory of Singapore, Nanyang Technological University, Singapore 639798*

---

[*]*Corresponding author e-mail address:* ajgarner@marine.rutgers.edu

## Abstract

The modern era of scientific global-mean sea-level rise (SLR) projections began in the early 1980s. In subsequent decades, understanding of driving processes has improved, and new methodologies have been developed. Nonetheless, despite more than 70 studies, future SLR remains deeply uncertain. To facilitate understanding of the historical development of SLR projections and contextualize current projections, we have compiled a comprehensive database of $21^{st}$ century global SLR projections. Although central estimates of $21^{st}$ century global-mean SLR have been relatively consistent, the range of projected SLR has varied greatly over time.. Among studies providing multiple estimates, the range of upper projections shrank from 1.3 – 1.8 m during the 1980s to 0.6 – 0.9 m in 2007, before expanding again to 0.5 – 2.5 m since 2013. Upper projections of SLR from individual studies are generally higher than upper projections from the Intergovernmental Panel on Climate Change, potentially due to differing percentile bounds, or a pre-disposition of consensus-based approaches toward relatively conservative outcomes.

## Plain Language Summary

In spite of more than 35 years of research, and over 70 individual studies, the upper bound of future global-mean sea-level rise (SLR) remains deeply uncertain. In an effort to improve understanding of the history of the science behind projected SLR, we present and analyze the first comprehensive database of $21^{st}$ century global-mean SLR projections. Results show a reduction in the range of SLR projections from the first studies through the mid-2000s that has since reversed. In addition, results from this work indicate a tendency for IPCC reports to "err on the side of least drama"—a conservative bias that could potentially impede risk management.

## 1. Introduction

Coastal populations and associated economic assets have increased steadily in recent decades (Neumann et al., 2015); by 2100, the population within 10 m elevation of mean sea level could exceed 830 million (Merkens et al., 2016). As coastal populations expand, the risks associated with sea-level rise (SLR) are also continuing to grow (P. U. Clark et al., 2016). Consequently, there is rapidly expanding demand for SLR projections at both global and local scales, but care is needed to ensure that these projections and their estimated uncertainties accurately reflect scientific knowledge (e.g., Sweet et al., 2017). An understanding of the historical evolution of sea-level projections provides crucial context for interpreting the current state of the art.

In the late 1970s and early 1980s, a growing awareness of the potential instability of the West Antarctic Ice Sheet (WAIS; e.g., J. A. Clark & Lingle, 1977) and the potential impact of global warming on sea level led to the development of the first modern projections of 21$^{st}$ century global mean SLR (Gornitz et al., 1982; Hoffman et al., 1983). These projections began with simple statistical models of the relationship between global mean sea level and temperature (Gornitz et al., 1982), but soon became dominated by approaches that aimed to assess likely future SLR by integrating model- and literature-based projections for individual processes (e.g., Hoffman et al., 1983). Policymakers recognized the need to incorporate these emerging projections into decision processes, leading to a National Research Council (U.S.) study (NRC, 1987) that developed a discrete set of scenarios, eventually adapted by the US Army Corps of Engineers (USACE, 1989). In subsequent years, understanding of processes driving SLR improved, and new scientific and analytic tools were developed. Thus, methods of projecting future SLR expanded to include process-based models (Raper et al., 1996), semi-empirical models (Rahmstorf, 2007), unstructured expert judgments (B. P. Horton et al., 2014), and probabilistic assessments (Kopp et al., 2014).

Despite methodological advances, the upper bound of sea-level projections remains deeply uncertain, with no single agreed-upon probability distribution, and no generally accepted "best" estimation method (Kopp et al., 2017). Although there have been attempts to summarize both the difficulties associated with projecting future SLR and the inevitable differences among SLR projections (Oppenheimer & Alley, 2016), to date there has been no attempt to develop a comprehensive database to examine the historical development of global-mean SLR projections.

Here, we have compiled a comprehensive database of studies from 1983 - 2018 that project future global-mean SLR at the end of the 21st century. It should be noted that there are a variety of factors that lead to differences in projected global-mean SLR across studies, including approaches to characterizing risk, specific SLR components included and analyzed in any given study, relative reliance upon global climate models compared to other sources of information, and assumptions about emissions scenarios and future climate forcing. Because of the diverse sets of assumptions and goals used by individual studies, it is often not possible to make direct comparisons between separate studies; however, we nonetheless attempt to illuminate and contextualize the varied sources of differences across SLR projections as a whole. As the number of publications on this topic continues to expand, this database may provide context for researchers and decision makers as they grapple with challenges from methodological choices to deep uncertainty.

## 2. Database of sea-level rise projections

The database (Table S1) includes SLR projections from 74 different studies (Fig. 1a), which are subdivided into eight methodological categories (Table 1, Fig. S1). The 21$^{st}$ century SLR projections in the database are also categorized by low, mid, and high emissions scenarios.

Table S2 shows the categorization of emission scenarios used in Intergovernmental Panel on Climate Change (IPCC) reports for this database (Church et al., 2001, 2013; Hartmann et al., 2013; Meehl et al., 2007; Rogelj et al., 2012; Warrick et al., 1996; Warrick & Oerlemans, 1990). SLR projections made under geoengineering scenarios are not included.

Where possible, the $5^{th}$, $50^{th}$, and $95^{th}$ percentile estimates from the original studies are used as lower, central, and upper estimates for each study-by-scenario in the database. However, this is not always possible, because 1) some studies use different definitions of lower, central, and upper estimates (for example, a $10^{th}$, $50^{th}$, and $90^{th}$ percentile, or a mean ± one standard deviation); and 2) not all studies provide a range of estimates, but instead report a single value This is particularly true for many of the early studies; in such cases, the values provided are considered central estimates. We also note that the $5^{th}$ to $95^{th}$ percentile range used in this analysis differs from the ranges used in some of the IPCC assessment reports. The first (FAR), second (SAR), and third (TAR) assessment reports provide extreme ranges of SLR across scenarios, the fourth assessment report (AR4) provides a span of the 5-95% range across scenarios, and the fifth assessment report (AR5) focuses on a central or "likely" (at least 66% probability) range of SLR across scenarios (Church et al., 2001, 2013; Meehl et al., 2007; Warrick et al., 1996; Warrick & Oerlemans, 1990; Table S3). The evolution of emissions scenarios, coupled with methodological choices, inevitably limits direct comparisons of how and why SLR projections have evolved over time.

The number of projections for each study in the database is often related to the number of different climate scenarios used. However, some studies (particularly probabilistic studies), have single projections comprised of thousands of additional SLR samples. For example, the database includes three projections from Kopp et al. (2014)—one each for RCP2.6, 4.5, and 8.5—but each of these projections was based upon 10,000 Monte Carlo samples of SLR (Kopp et al., 2014).

Each study in the database includes the following fields: 1) year in which the study was published; 2) lead author of the study; 3) methodological approach; 4) base year(s) for the projections; 5) end year(s) for the projections; 6) emissions scenario used; 7) emissions scenario category (low, mid, or high); 8) lower estimate of sea-level change; 9) lower rate of sea-level change; 10) definition of lower estimate of sea-level change; 11) central estimate of sea-level change; 12) central rate of sea-level change; 13) definition of central estimate of sea-level change; 14) upper estimate of sea-level change; 15) upper rate of sea-level change; 16) definition of upper estimate of sea-level change. Not all of these fields are available for each SLR projection; for example, some studies include only a central estimate, rather than a lower, central, and upper estimate of SLR. Note that lower, central, and upper estimates should not be confused with low, mid, and high emissions scenarios. For example, a study that provides a single upper estimate of SLR based on a high emissions scenario would be classified as a central estimate with a scenario type classified as "High".

The database does not include studies that looked at just one or two components of global SLR, but rather includes only studies that have at least in some way incorporated 1) thermal expansion, 2) polar ice sheets, and 3) glaciers and ice caps. Although we have attempted to include all projections of 21$^{st}$ century SLR, it is perhaps inevitable that we have missed a small number of projections that should have been included.

**2.1. Projection Windows**

Projection windows for the SLR projections included in the database are determined by the base year(s) and end year(s) used by each individual study, and are not uniform across different studies. Base years for entries tend to vary with the time at which each projection was made, but, when analyzing 21$^{st}$ century SLR estimates, we have required that end years for studies extend to at least the final decade of the 21$^{st}$ century. So, for example, a study with an end year of 2080 would not be included in such analysis, but, a study with an end year window spanning 2070-2099 would be included. We have not used these same requirements in analyzing evolving methodologies for SLR studies (e.g., Fig. 1b); instead, we have included all relevant unique SLR projections as we consider how this aspect of the history of the science has evolved over time.

In order to generate consistency across studies and create a common framework in which to compare different SLR projections, we have normalized the sea-level estimates by using the base and end years to calculate average rates of SLR for each projection in the database, as follows:

$$SLR_{Adj} = SLR\left(\frac{100}{(Y-Y_0)}\right), \qquad [1]$$

where $SLR_{Adj}$ is the normalized SLR projection (the rate of sea-level change), $SLR$ is the SLR reported in the original study, $Y$ is the study end year, and $Y_0$ is the study baseline year. In cases where a range of years is used for either the study end point, or for the study baseline, we use the central year from the range for Eq. [1] above. This normalization process results in little change to the overall values of SLR at the end of the 21$^{st}$ century that we report here compared to values given in the original studies, given that most projection windows are already close to 100 years. We do note that, because of inter-annual and decadal variations in SLR, and because of the acceleration of most projections, this normalization process may slightly bias some results compared to others; however, this approach is nonetheless useful in allowing us to standardize the different projections for easier comparison across studies.

## 3. Evolution of sea-level rise estimates and ranges

SLR projections prior to the first IPCC report (between 1982 and 1990) included the first semi-empirical study, which projected global mean sea level at 2050 (Gornitz et al., 1982), as well as the first model hybrid study (Hoffman et al., 1983). However, most of the projections from this time period used a literature synthesis approach to estimate future SLR (e.g., Thomas, 1987; Fig. 1b, 2, S2). In total, there were only 16 published projections from 1982 to 1989 (Fig. 1a). SLR projections from this time period have the greatest range of any time period across the 36 years which the database spans (Fig. 2, S3). Projections of 2100 sea level range from -1.0 m for a scenario of drastically reduced greenhouse gas emissions relative to 1985 and low climate sensitivity (W. C. Clark et al., 1988) to 3.1 m for a scenario that included 4.0 °C warming in response to a doubling of $CO_2$ concentrations (Hoffman et al., 1986).

The range of these projections may reflect gaps in scientific knowledge about the processes that contribute to SLR, reflected in assumptions used to produce projections. For example, Hoffman et al. (1983) noted the problem of determining population and productivity growth, atmospheric and climatic change, and oceanic and glacial response. They also remarked that differences in estimates of SLR were due to insufficient scientific understanding and deficiencies in the methods used for constructing estimates, before stating that these shortcomings could be overcome with future research (Hoffman et al., 1983).

IPCC FAR, published in 1990, noted the difficulty in comparing future SLR values from different studies with varying time periods (end-years between 2025 and 2100), and differing assumptions. FAR generated global SLR projections of 0.31 to 1.1 m (extreme range of all 4 IPCC scenarios), based on IPCC FAR greenhouse gas forcing scenarios (Warrick & Oerlemans, 1990; Table S3). The major contributions to SLR in FAR projections were thermal expansion and glaciers and small ice caps (Warrick & Oerlemans, 1990). It was assumed that the major ice sheets would remain stable throughout the $21^{st}$ century, with only small contributions to SLR associated with changes in surface mass balance (Warrick & Oerlemans, 1990). FAR SLR projections included a minor positive contribution from the Greenland Ice Sheet, and a minor negative contribution from ice mass gains in Antarctica (Warrick & Oerlemans, 1990).

SLR projections made between IPCC FAR and SAR reports (1991 and 1995) included model synthesis studies (e.g., Wigley & Raper, 1993), as well as the first probabilistic study in the database (Titus & Narayanan, 1995). Projections of $21^{st}$ century SLR ranged from -0.26 m (the $2.5^{th}$ percentile from a probability distribution based on the IS92A-F scenarios; Titus & Narayanan, 1995), to 1.13 m (for the IPCC BAU scenario; Wigley & Raper, 1993).

SAR was published in 1996 and drew upon projections published in FAR (Warrick & Oerlemans, 1990), as well as the new projections. However, as with FAR, SAR noted the difficulty of comparing previous studies due to their varying assumptions related to emission scenarios, greenhouse gas concentrations, radiative forcing, and climate sensitivity. As a synthesis of the published studies to date, SAR provided a set of projections using the IPCC emission scenarios that were slightly lower than those from FAR, ranging from 0.13 to 0.94 m (Table S3), mainly due to lower global temperature projections (Warrick et al., 1996).

SLR projections between SAR and TAR reports of the IPCC (1996 and 2001) included a number of projections from model synthesis studies (de Wolde et al., 1997). Projections of SLR at the end of the 21$^{st}$ century from studies during this time period ranged from 0.07 m for a low scenario where $CO_2$ concentration stabilizes at 450 ppmv and low ice melt parameter values are used, to 2.9 m for a high scenario where $CO_2$ concentration stabilizes at 650 ppmv and high ice melt parameter values are used (Raper et al., 1996).

TAR drew upon some of the projections that are found in the database between 1996 and 2001 (Raper et al., 1996; de Wolde et al., 1997) but primarily focused on new model synthesis projections using Atmosphere-Ocean Global Climate Models (AOGCMs). The range of these 21$^{st}$ century global SLR projections extended from 0.09 to 0.88 m (Church et al., 2001) across the 35 SRES scenarios (Table S3). Projections for thermal expansion were based on a simple climate model (Raper et al., 1996), ice sheet mass balance sensitivities were derived from AOGCMs, and ice-dynamical changes in the WAIS were not included, as it was generally believed that major contributions to SLR due to loss of grounded ice from the WAIS was very unlikely during the 21$^{st}$ century (Church et al., 2001).

Between the publication of TAR in 2001 and AR4 in 2007, there were no new projections of global SLR, although there were numerous publications exploring the mechanisms that drive SLR (Gregory et al., 2001; Levermann et al., 2005; Oerlemans, 2001; Suzuki et al., 2005). These included studies related to thermal expansion (Gregory et al., 2001), ocean density and circulation changes (Gregory et al., 2001; Levermann et al., 2005), glaciers (Oerlemans, 2001), and the Greenland and Antarctic Ice Sheets (Suzuki et al., 2005). AR4 authors drew upon this literature in the development of their projections, which ranged from 0.18 to 0.59 m (Meehl et al., 2007). This range was notably lower than the TAR range, primarily because it did not account for contributions from Greenland glaciers and West Antarctic ice streams (Meehl et al., 2007). AR4 projections included a large contribution from thermal expansion, with additional positive contributions from glaciers, ice caps, and Greenland via surface mass balance, through negative contributions from a snowier Antarctic Ice Sheet. AR4 authors noted that much uncertainty remained about ice flow in Greenland glaciers and West Antarctica, and that although the primary AR4 projections did not account for such contributions, increased ice discharge from these processes could greatly increase future SLR (Meehl et al., 2007). The discussion of future SLR in AR4 indicated a need for more research on the subject of future polar ice sheet response to continued global warming.

Dissatisfaction with physical models of SLR (Rahmstorf, 2007), along with growing observational evidence of ice sheet loss (e.g., Rignot et al., 2011) helped spur a significant increase in the number of SLR projections (20 new studies) between the publication of AR4 in 2007 and the publication of AR5 in 2013. New projections were dominated by the renaissance of semi-empirical models (Rahmstorf, 2007; Fig. 1b, 2, S2). Rahmstorf (2007) suggested the historical relationship between global mean surface temperature and rate of sea level change, combined with projections of global mean surface temperature, could yield improved SLR projections relative to those based on physical modeling. Between 2007 and 2013, the range of SLR for 2100 from semi-empirical models was 0.17 m to 2.05 m. These projections are, however, limited by the structural uncertainty regarding whether empirical connections observed during the instrumental or proxy time periods will remain unchanged in the future, and are also sensitive to the choice of data used for calibration (Rahmstorf et al., 2012).

AR5 authors drew upon results from semi-empirical models (e.g., R. Horton et al., 2008; Rahmstorf, 2007), but assigned these projections low confidence, while also drawing upon various model synthesis and model hybrid studies, to which they assigned greater confidence (e.g., Sriver et al., 2012). AR5 provided their own projections of $21^{st}$ century SLR from process-based models, with a likely (at least 66% probability) range of 0.26 – 0.82 m (Table S3). This range, although comparable to the range given in TAR, represented a significant upward revision from the values reported in AR4, primarily due to the inclusion of more rapid changes in Greenland and Antarctic ice sheets. However, AR5 also noted that additional SLR up to several tenths of a meter was possible due to Marine Ice Sheet Instability (MISI), a process that was not included in the estimate of Antarctic ice-sheet rapid dynamics due to imprecise estimates of the likelihood of such a contribution.

Twenty-eight studies and more than 90 projections (> 30% of the total number of SLR projections in the database) have been published from 2013 to the present. This time period has also seen a proliferation of national and subnational sea-level assessment documents (Hall et al., n.d.). The range of 2100 SLR across these projections is 0.16 m to 2.54 m, which is both broader and higher compared to projections made between TAR and AR5 (Fig. 2, S3, 3). The change in range reflects increased uncertainty about maximum contributions of the Greenland and Antarctic Ice Sheets to SLR (DeConto & Pollard, 2016; Levermann et al., 2013).

Although all of the categories of SLR projections have been represented during this recent time period (Fig. 1b), a major new development since AR5 has been the spread of probabilistic methodologies (e.g., Kopp et al., 2014, 2017) and the introduction of projections derived from expert judgement methodologies (Bamber & Aspinall, 2013; B. P. Horton et al., 2014). The development of probabilistic methodologies and utilization of structured expert judgement methodologies (Fig. 1b) support exploration of extreme SLR possibilities, which can generate the greatest risks, and thus play an important role in coastal risk management and planning

(Kopp et al., 2014). Although a few earlier assessments involving decision makers attempted to provide upper-bound SLR projections for risk-based decision contexts (R. Horton et al., 2010), structured expert judgment and probabilistic approaches hold promise for mainstreaming consideration of high-end outcomes via decision-maker engagement and co-production of knowledge (R. Horton et al., 2015; Sweet et al., 2017). Such projections address the inadequacy of presenting only central ranges for SLR projections, as the likely (at least 66% probability) ranges provide no information about the highest 17% of outcomes (Kopp et al., 2014). However, while probabilistic methodologies represent an important addition to SLR projection methods, large uncertainties remain about key processes influencing individual SLR components, how different components may interact in a changing climate, and future concentrations of radiatively important agents and associated climate sensitivity.

## 4. IPCC Sea-Level Rise Projections: Erring on the side of least drama?

AR5 projected a 'likely' (i.e., at least 66% probability) global-mean SLR of 0.52-0.98 m in the case of unmitigated growth of emissions (RCP8.5) by 2100, relative to 1986-2005 (Church et al., 2013). However, many projections for high emissions scenarios from individual studies (Fig. 2a, S2a, 4) are much greater than 1 m. This trend has been particularly true for upper estimates of SLR from high-emissions scenarios (Fig. 2, 4), with the majority of these projections exceeding the upper estimates provided by the IPCC assessment reports. This result aligns with the findings of Horton et al. (2014), in which the authors noted that most experts predicted greater amounts of SLR by 2100 than the 'likely' range of 21$^{st}$ century SLR given in AR5 (Church et al., 2013).

Although the IPCC acknowledges its limitations in projecting future SLR (Church et al., 2001, 2013; Meehl et al., 2007; Warrick et al., 1996; Warrick & Oerlemans, 1990), caveat language included in the reports tends to get filtered out in headline numbers. There are several reasons that projected SLR from the IPCC reports may tend to be lower than upper estimates from other studies. First, the type of model-based studies on which AR5 placed the greatest emphasis may be relatively insensitive to potential changes in ice sheet behavior as temperatures rise (Church et al., 2013). Second, the IPCC percentile bounds may be narrower than other studies use to project ranges of SLR. For example, AR5 focused on a 'likely' (approximately 17$^{th}$ to 83$^{rd}$ percentile) range of projected SLR, and did not attempt to provide quantitative information about less likely outcomes. Third, consensus-based approaches like the IPCC, with their large number of authors, may be predisposed to relatively conservative outcomes—both in the overall assessment of the literature and through communication choices, such as which percentiles to emphasize (Brysse et al., 2013). Finally, the IPCC knowledge development process only includes scientists. Without the inclusion of decision makers who manage coastal risk in the development of that knowledge, the utility of the IPCC for planning and managing coastal risk, especially at regional to local scales, is hard to gauge. Of the small number of SLR projections that have included participation and input from decisions makers, all have considered high-end estimates as useful for considering impacts and consequences of SLR, particularly examining assets for which we can tolerate only a low probability of hazard occurrence, due to large consequences should the hazard occur (e.g., nuclear power plants or other energy infrastructure).

Ultimately, the IPCC reports have tended to err on the side of providing intentionally cautious and conservative estimates of SLR, rather than focusing on less likely, extreme possibilities that would be of high consequence, should they occur. This bias towards such cautious estimates has been described previously as "erring on the side of least drama" (Brysse et al., 2013). Many individual studies, both globally and locally (R. M. Horton et al., 2011), have not constrained the ranges of their SLR projections in the same conservative manner as the IPCC reports. Rather than erring on the side of least drama (Brysse et al., 2013), such studies better encompass less likely, but more severe outcomes of future SLR that may be of greater interest to audiences concerned with risk-based perspectives (e.g., Rosenzweig et al., 2014).

This database documents the development of a 36-year-old body of scientific knowledge. Throughout this history, the IPCC remains a useful foil. Gradually, over the latter reports (TAR, AR4, and AR5), IPCC has become a judge of the standard of scientific practice, deeming certain methods (e.g., physical models) credible and others perhaps not yet so (e.g., semi-empirical models; Fig. 4). The conservative bias exhibited by IPCC analyses may in part be due to IPCC Working Group 1's development of knowledge solely within the epistemic domain of the natural sciences (e.g., McNie et al., 2016).

Scientists evaluating science can lead to "cracks of bias" in many fields (Sarewitz, 2012). The IPCC is designed to influence the United Nations Framework on Climate Change Convention (UNFCCC), which is critically important for curbing global emissions and, by inference, SLR. However, if the bias toward a lower, central range is due to epistemic norms, it suggests that the science-policy interface between the IPCC and UNFCCC or other decision-making bodies may be too limited to allow for appropriate participation from decision-makers and the development of useful knowledge for climate adaptation (e.g., Parris et al., 2015).

## 5. Uncertainty characterization in recent SLR projections on different time scales

The comparison of SLR projections has historically been challenging, due to projections' varying timescales, inconsistent assumptions about emissions, radiative forcing, and climate sensitivities, and ambiguously defined lower, central, and upper estimates of SLR. However, the broad use of RCP scenarios and the adoption of explicit Bayesian probabilities (not only in probabilistic projections, but also in semi-empirical projections and model syntheses) across many of the SLR projections made since AR5 has helped to eliminate ambiguity at least in how emission scenarios and lower and upper estimates of SLR are defined (e.g., Grinsted et al., 2015; Kopp et al., 2014, 2017).

As discussed in Kopp et al. (2017), upper bounds of future SLR projections remain deeply uncertain. Deep uncertainty has been defined as "the condition in which analysts do not know or the parties to a decision cannot agree upon 1) the appropriate models to describe interactions among a system's variables; 2) the probability distributions to represent uncertainty about key parameters in the models; and/or 3) how to value the desirability of alternative outcomes" (Lempert et al., 2003). The deeply uncertain nature of SLR projections is evident by the fact that there is no unique probability distribution of future sea-level; thus, it is unlikely that there will be any particular method that is found to be best for estimating future sea-level change anytime in the near future (Kopp et al., 2017). Therefore, it is useful to compare multiple possible SLR distributions (Fig. 5).

While there is significant spread in SLR projections for the end of the 21$^{st}$ century, the same is not necessarily true of SLR projections on shorter time scales. We have compared the partial cumulative distribution functions (CDFs) based on the selected values reported in semi-empirical and probabilistic studies since AR5 (Fig. 5). There is far greater agreement among studies about SLR in 2050 compared to 2100, although methodology appears to be more important than RCP for 2050 projections, whereas 2100 projections appear to be strongly influenced by RCP (Fig. 5). The overall spread of projections is far more constrained for 2050 projections (5$^{th}$ percentile of 0.12 to 0.25 m, 95$^{th}$ percentile of 0.21 to 0.48 m; Table S4 than for 2100 (5$^{th}$ percentile of 0.21 to 1.09 m, 95$^{th}$ percentile of 0.53 to 2.43 m; Table S4. These results emphasize the deep uncertainty that scientists face in trying to predict the contributions to SLR at 2100 from various components, especially ice sheets, compared to the more tangible contributions to SLR on shorter time scales.

The majority of studies seeking to project future SLR have focused on the year 2100. However, as the world moves closer to the year 2100, it is essential to understand SLR and the impacts of rising sea-levels on longer time scales (Brown et al., 2018; P. U. Clark et al., 2016; Levermann et al., 2013). A few recent studies have sought to project SLR for 2300, with median estimates of global-mean SLR ranging from 1.00 m under RCP2.6 to 11.69 m under RCP8.5 (Brown et al., 2018; Kopp et al., 2014, 2017; Nauels, Meinshausen, et al., 2017; Schaeffer et al., 2012), while studies looking at multi-millennial sea-level commitments have suggested over 20 m of future global-mean SLR for emissions scenarios similar to RCP4.5 (P. U. Clark et al., 2016; Levermann et al., 2013).

## 6. Conclusion

In 1983, Hoffman et al. (1983) issued a call for further investigation of the components that contribute SLR, suggesting that with further research, differences in estimates of future SLR due to inadequate scientific knowledge and shortcomings in the methods used to construct estimates could be overcome, allowing for more precise estimates of future changes in sea level. More than a generation later, future SLR remains deeply uncertain in nature, in spite of more than 70 unique studies projecting future SLR, and additional studies investigating individual components of SLR, as well as significant developments in methodological approaches.

This database illustrates the many ways in which methodologies of SLR have evolved over the last four decades. From projections made during the 1980s prior to FAR to the publication of AR4, there was ultimately a narrowing and a lowering of the range of projected 21$^{st}$ century SLR (from 1.32 – 1.81 m to 0.57 – 0.86 m for upper projections, and from 0.43 – 1.20 m to 0.09 – 0.18 m for lower projections; Fig. 3) across the studies in the database (Figs. 2, 3, S2, S3, and S4). Since AR4, however, the range of SLR projections among individual studies has increased, with a range of 0.46 – 2.54 m for upper projections and a range of 0.16 to 1.55 m for lower projections published since AR5 (Fig. 3).

The narrowing of SLR projections from the 1980s until AR4, followed by the broadening of this range since AR4, may be an example of the phenomenon of "negative learning", or the departure over time of scientific beliefs from the prior answer due to the introduction of new technical information (Oppenheimer et al., 2008). For example, in the specific case of SLR projections, it is possible that the narrowing of projections prior to AR4 in the period immediately prior to observed changes in ice sheet behavior was somewhat premature, a trend that has now begun to be reversed. In climate science, this phenomenon can often lead to confusion for decision makers and policy makers, though waiting for positive learning (often characterized by observations leading the models) can result in costly consequences (Oppenheimer et al., 2008). As new rounds of SLR projections are developed, a better awareness and understanding of the history of the science could be beneficial—highlighting the importance of a database such as the one developed here. In the future, coordinated programs and agreement on standardized approaches could facilitate efforts to make comparisons that illuminate all the reasons why projections differ across studies, something that is not possible given the diverse methods and impossibility of modifying many of the studies to date. As awareness grows that other aspects of the climate system may be characterized by deep uncertainty as well (e.g., Lenton et al., 2008), examples of how the SLR and coastal risk communities have integrated different types of information and projection approaches over time may prove instructive.

## Acknowledgements

AJG was supported by National Science Foundation EAR Postdoctoral Fellowship 1625150, the Community Foundation of New Jersey, and David and Arlene McGlade. REK was supported in part by National Science Foundation grant ICER-1663807 and NASA grant 80NSSC17K0698. BPH is supported by the National Research Foundation Singapore and the Singapore Ministry of Education under the Research Centres of Excellence initiative and National Science Foundation OCE-1458904. This paper is a contribution to International Geoscience Programme (IGCP) Project 639, "Sea Level Change from Minutes to Millennia." This is Earth Observatory of Singapore contribution XX. Data used for this paper is available in the Supporting Information and will be maintained in the following github repository: [insert URL here upon acceptance].

## Author Contributions

AJG, JLW, AP, REK, RMH, JTO, and BPH designed research; AJG, JLW, AP, REK, RMH, and BPH performed research; AJG performed analysis; AJG, JLW, AP, REK, RMH, JTO, and BPH wrote the paper.

**Figures**

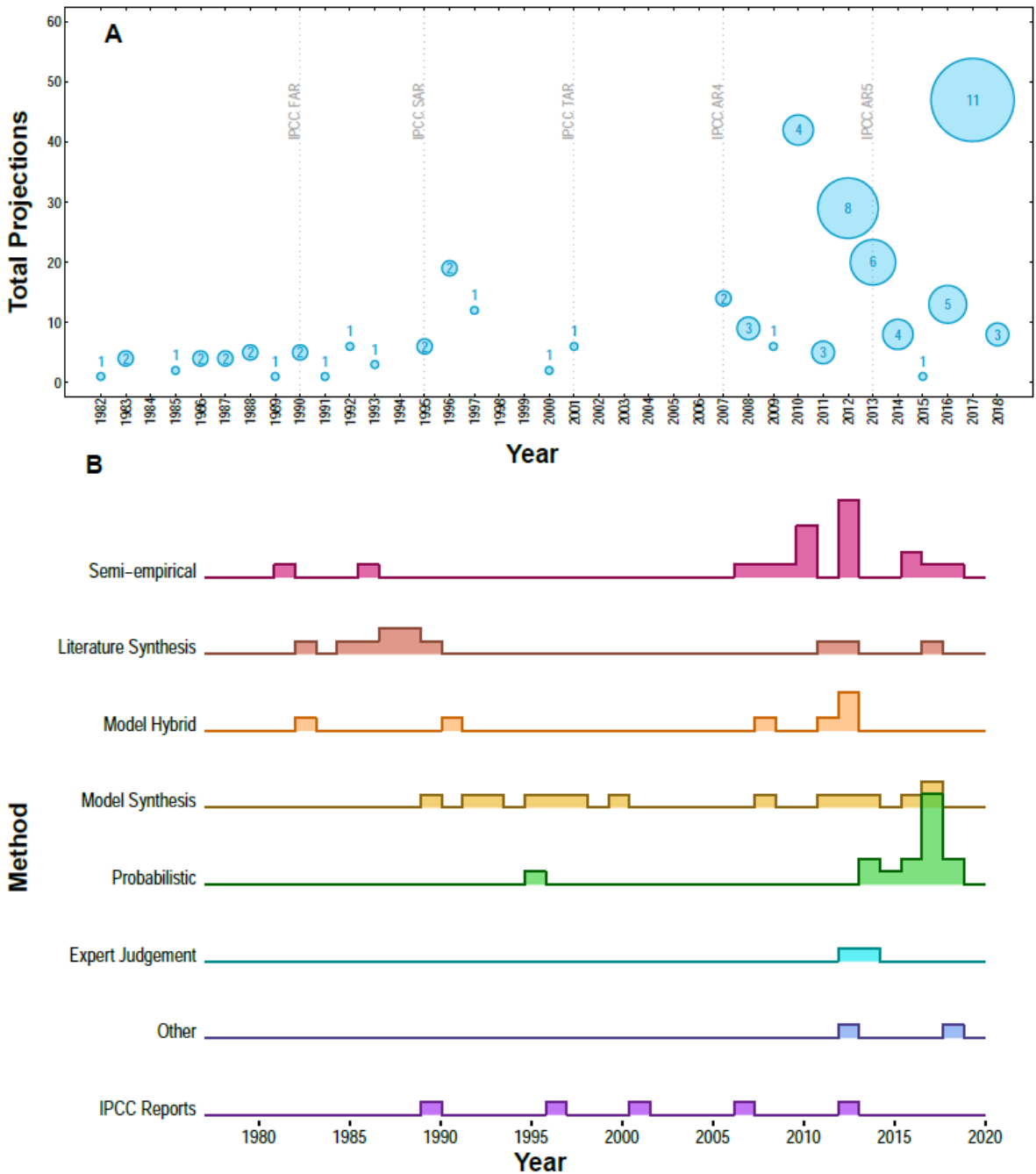

**Figure 1:** Total projections and methodology time series of 21st century SLR projections. (a) Total number of 21st century SLR projections per study year, where the number of individual studies producing projections each year is indicated by size and numbers in blue for each point. Many studies produce multiple projections, including different projections for different emissions scenarios. The year in which the study was published is shown on the x-axis. Gray dashed lines indicate years of IPCC reports. (b) Density time series of relative number of studies for each methodology category published from 1982 to the present.

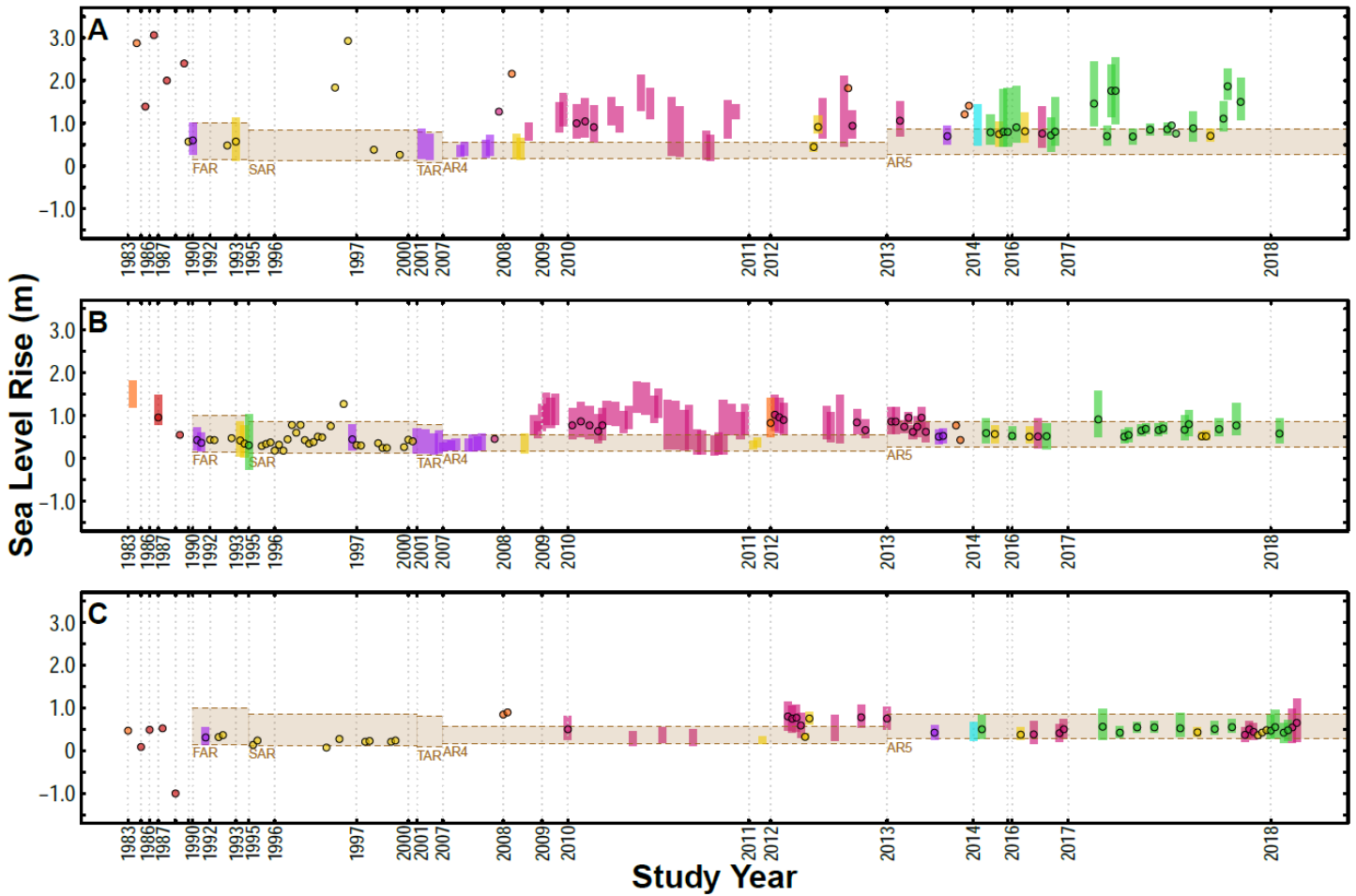

Figure 2: Evolution of the ranges of SLR projections from 1983 – 2018. Circular points represent central SLR projections; bars extend from the lower SLR projection to the upper SLR projection for (a) high emissions scenarios, (b) middle emissions scenarios, and (c) low emissions scenarios. Where possible, bars show the $5^{th} - 95^{th}$ percentile range of individual projections. Bar and point colors correspond to the methodology used by each study, and are as in Fig. 1b: semi-empirical (pink), literature synthesis (red), model hybrid (orange), model synthesis (yellow), probabilistic (green), expert judgement (cyan), other (blue), and IPCC reports (purple). Tan shaded regions and dashed lines represent the ranges of SLR from the IPCC reports, as in Table S3: the extreme range of projections for IPCC FAR and SAR, the range of all AOGCMs and SRES scenarios for TAR, the 5-95% range across SRES scenarios for AR4 (which do not include dynamic ice sheet response), and the 'likely' ($17^{th} - 83^{rd}$ percentile) range from process-based models for AR5 (potential rise above this range as specified in AR5 is not included in the shaded region). Note that 1) time steps are non-uniform, in order to clearly show all projections, 2) a small number of projections in the database have no specified emissions scenario, and are left off of this figure, and 3) projections have been normalized using Eq. [1] as specified in Section 2.1.

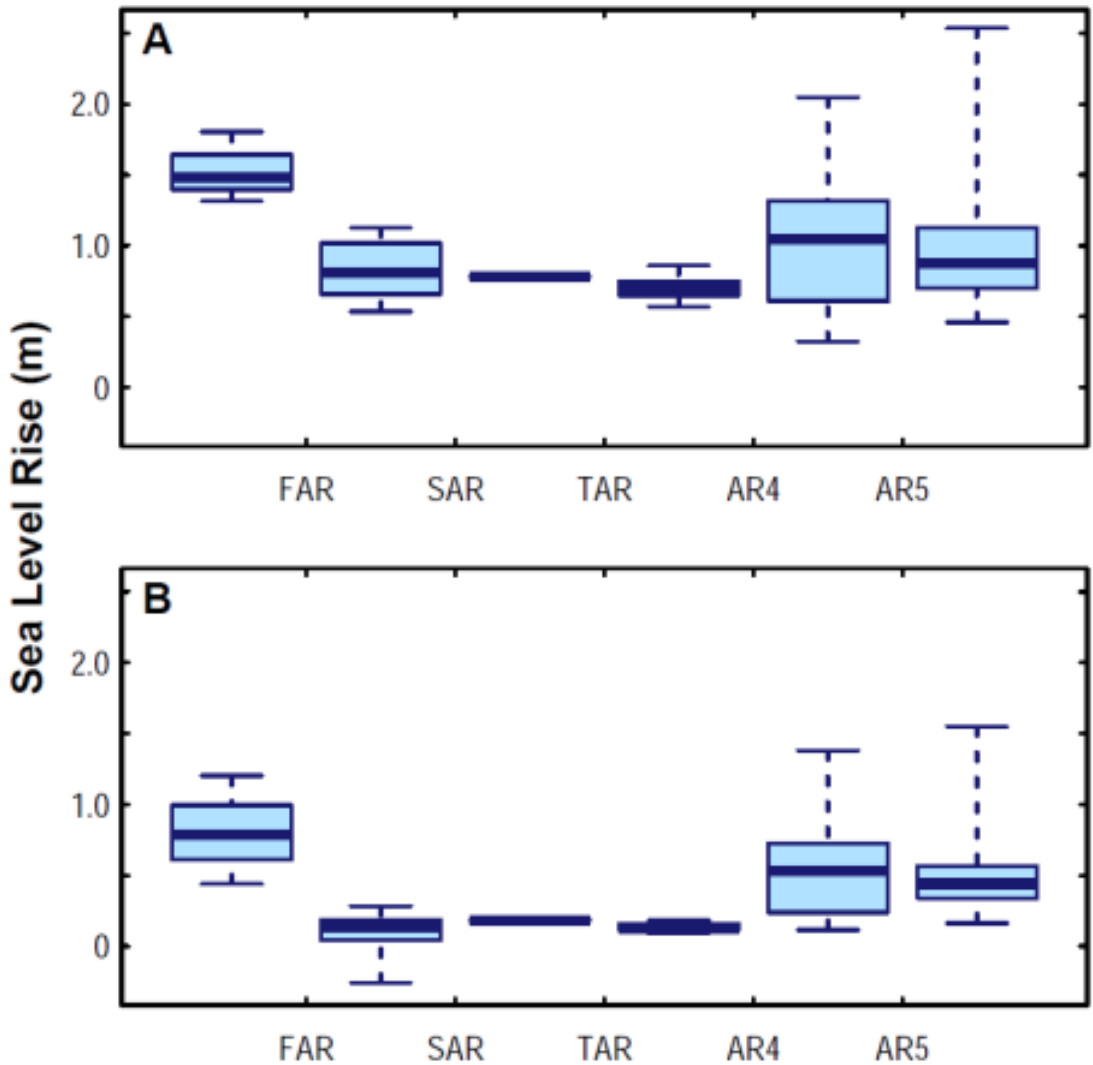

**Figure 3:** Box and whisker plots showing SLR ranges over time. Shown are the varying ranges of (a) upper SLR projections and (b) lower SLR projections. Box edges extend from the 25th to 75th percentiles; the solid line in each box shows the 50th percentile. Whiskers extend to data extremes, essentially ranging from 0 to 100th percentiles to show the full range of SLR projections in each case. The horizontal axis uses the Intergovernmental Panel on Climate Change assessment reports to divide the literature based on publication date.

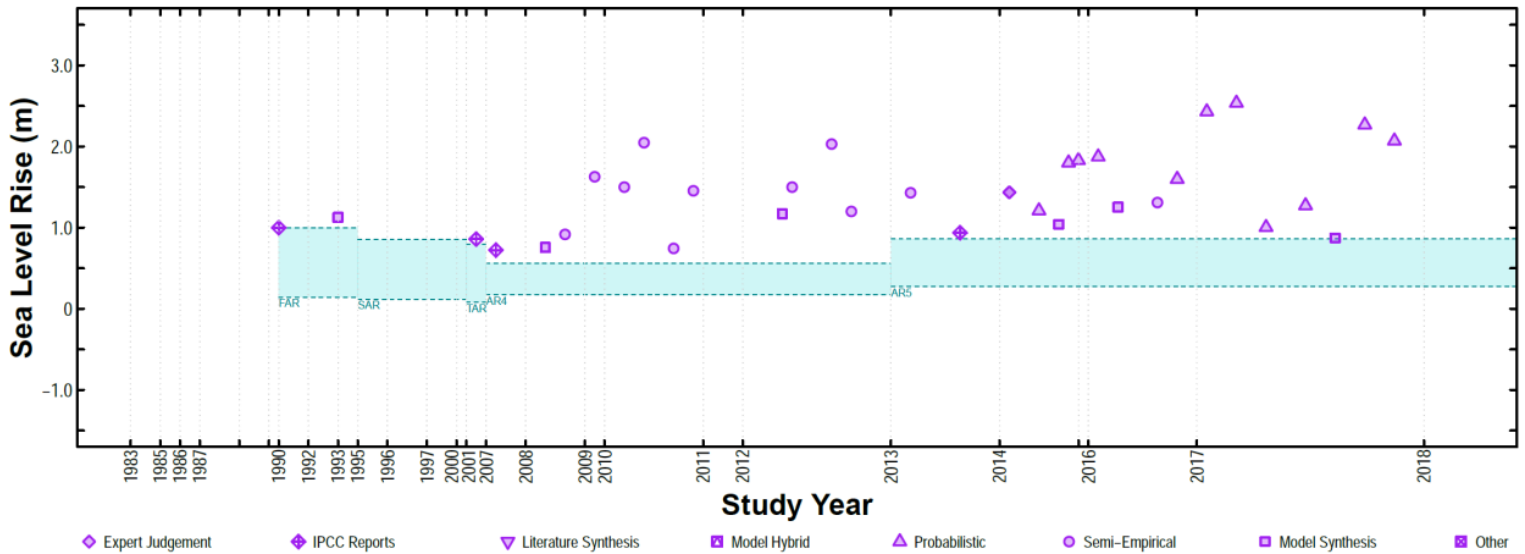

**Figure 4:** Comparison of upper estimates for high emissions scenarios from individual studies to IPCC projected ranges of SLR. Shown are the upper estimates of SLR for high emissions scenarios from 1983-2018 (purple), and the IPCC projected SLR ranges (blue). Where possible, upper estimates from high emissions scenarios show the 95$^{th}$ percentile estimate; ranges for IPCC reports are as shown in Table S3: the extreme range of projections for IPCC FAR and SAR, the range of all AOGCMs and SRES scenarios for TAR, the 5-95% range across SRES scenarios for AR4 (which do not include dynamic ice sheet response), and the 'likely' (17$^{th}$ – 83$^{rd}$ percentile) range from process-based models for AR5 (potential rise above this range as specified in AR5 is not included in the shaded region). Note that time steps are non-uniform, in order to clearly show all projections, and projections have been normalized using Eq. [1] as specified in Section 2.1.

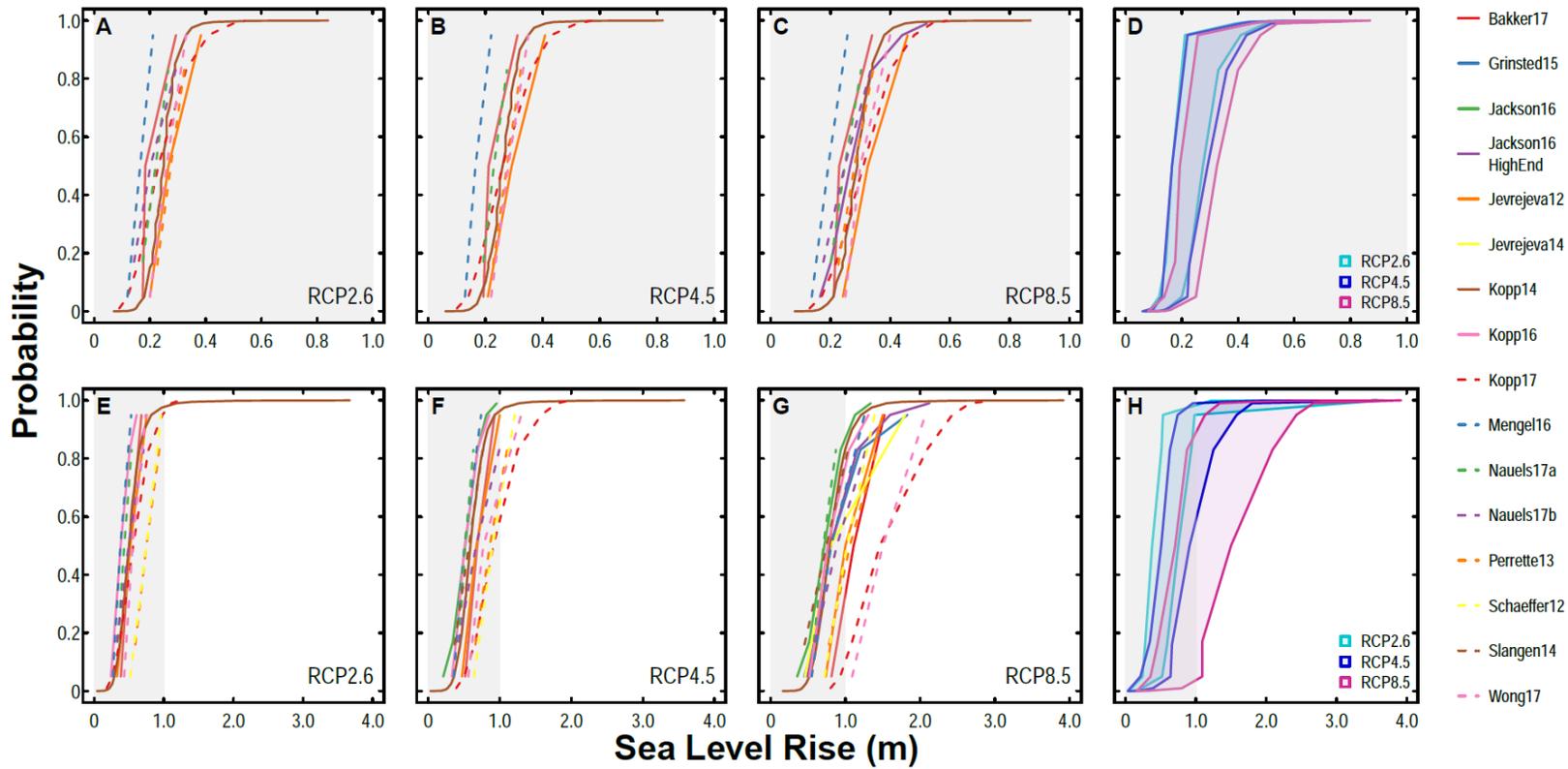

**Figure 5**: CDFs based on projections from semi-empirical, probabilistic, and model synthesis studies produced since AR5 for both 2050 (a-d) and 2100 (e-h). The right-most panel in each row shows regions representing the upper and lower bounds of CDFs for RCP2.6, RCP4.5, and RCP8.5 emissions scenarios.

**Tables**

Table 1 | Method categories of SLR projections published since 1982

| Method | Definition | Total Studies[*] |
|---|---|---|
| Semi-empirical models | *Calculations are based on a historical statistical relationship between global mean sea level and some other driving factor, such as temperature* | 19 |
| Literature Syntheses | *Previously published projections are used to estimate contributions from every individual component of SLR considered and generate a new future SLR projection* | 10 |
| Model Hybrid Studies | *A combination of physical and statistical modeling techniques are used to produce estimates for some, but not all, individual SLR components; other methods, such as literature syntheses, are used to estimate contributions from remaining SLR components* | 6 |
| Model Syntheses | *Some combination of physical and statistical modeling is used to estimate contributions from every individual component of SLR considered and generate a new future SLR projection* | 14 |
| Probabilistic Projections | *Different sub-models or lines of evidence are combined in a fashion intended to produce comprehensive probability distributions of future sea-level change* | 15 |
| Expert Judgement Studies | *Projections based on responses to broad surveys of experts active in the field of sea level* | 2 |
| Other Methods | *includes a small number of studies such as those that employed a combination of tide gauge data, satellite altimeter data, and data about the equilibrium sea level response to a warming climate* | 2 |
| IPCC Reports | *Projections provided in the IPCC Assessment reports* | 5 |

[*] Note that here, as in Fig. 1, we focus on the broad patterns of study methodologies over the course of the history of the science, and include all published SLR projections rather than limiting the total studies column to only the 21st century projections that are included in the SLR database.

**Supporting Information**

**For**

**"Evolution of 21$^{st}$ Century Sea-level Rise Projections"**

**Table S1 | Database of 21st Century SLR Projections**

| Lead Author | Method | Base Year(s) | End Year(s) | Emission Scenario | Scen. Cat. | Lower Est. (m) | Lower Rate (m/cen) | Lower Estimate Definition | Central Est. (m) | Central Rate (m/cen) | Central Estimate Definition | Upper Est. (m) | Upper Rate (m/cen) | Upper Estimate Definition |
|---|---|---|---|---|---|---|---|---|---|---|---|---|---|---|
| **1982** | | | | | | | | | | | | | | |
| Gornitz (Gornitz et al., 1982) | Semi-empirical | 1980 | 2050 | | None | 0.4 | 0.57 | Extrapolation of linear trend | | | | 0.6 | 0.86 | Extrapolation of linear trend |
| **1983** | | | | | | | | | | | | | | |
| Hoffman (Hoffman et al., 1983) | Model Hybrid | 1980 | 2100 | 1.5°C warming in response to 2xCO2 | Low | | | | 0.562 | 0.47 | Sum of component estimates | | | Sum of component estimates |
| | | | | 3.0°C warming in response to 2xCO2 | Mid | 1.444 | 1.2 | Sum of component estimates | | | | 2.166 | 1.8 | |
| | | | | 4.5°C warming in response to 2xCO2 | High | | | | 3.45 | 2.9 | | | | |
| Revelle (Revelle, 1983) | Literature Synthesis | 1980 | 2080 | | None | 0.52 | 0.52 | -25% of central estimate | 0.7 | 0.7 | Sum of component estimates | 0.88 | 0.88 | +25% of central estimate |
| **1985** | | | | | | | | | | | | | | |
| Polar Research Board (National Research Council, 1985) | Literature Synthesis | 1985 | 2100 | 2°C warming in response to 2xCO2 | Low | | | | 0.1 | 0.087 | Sum of component estimates | | | |
| | | | | 4°C warming in response to 2xCO2 | High | | | | 1.6 | 1.4 | | | | |
| **1986** | | | | | | | | | | | | | | |
| Hoffman (Hoffman et al., 1986) | Literature Synthesis | 1980 | 2100 | 2°C warming in response to 2xCO2 | Low | | | | 0.58655 | 0.49 | Sum of individual component contribution to SLR, Mid-value diffusivity | | | |
| | | | | 4°C warming in response to 2xCO2 | High | | | | 3.67337 | 3.1 | | | | |
| Robin (Robin, 1986) | Semi-empirical | 1980 | 2080 | 3.5°C ± 2.0°C warming in response to 2xCO2 | Mid | 0.24 | 0.24 | 1.5°C warming, 16 cm/°C SLR rate over 100 years | 0.56 | 0.56 | 3.0°C warming, 16 cm/°C SLR rate over 100 years | 0.88 | 0.88 | 4.5°C warming, 16 cm/°C SLR rate over 100 years |
| | | | | 3.5°C ± 2.0°C warming in response to 2xCO2 | Mid | 0.45 | 0.45 | 1.5°C warming, 30 cm/°C SLR rate over 100 years | 1.05 | 1.1 | 3.0°C warming, 30 cm/°C SLR rate over 100 years | 1.65 | 1.7 | 4.5°C warming, 30 cm/°C SLR rate over 100 years |
| **1987** | | | | | | | | | | | | | | |
| National Research Council (NRC, 1987) | Literature Synthesis | 1986 | 2100 | | None | 0.5 | 0.44 | | 1 | 0.88 | | 1.5 | 1.3 | |
| Thomas (Thomas, 1987) | Literature Synthesis | 1985 | 2100 | 3°C warming by 2050, constant global T thereafter | Mid | 0.9 | 0.78 | Sum of component estimates, Lower bound | 1.1 | 0.96 | Sum of component estimates, preferred values | 1.7 | 1.5 | Sum of component estimates, upper bound |
| | | | | 3°C warming by 2150 | Low | | | | 0.6 | 0.52 | | | | |
| | | | | 4.5°C warming in response to 2xCO2 | High | | | | 2.3 | 2.0 | | | | |

| Lead Author | Method | Base Year(s) | End Year(s) | Emission Scenario | Scen. Cat. | Lower Est. (m) | Lower Rate (m/cen) | Lower Estimate Definition | Central Est. (m) | Central Rate (m/cen) | Central Estimate Definition | Upper Est. (m) | Upper Rate (m/cen) | Upper Estimate Definition |
|---|---|---|---|---|---|---|---|---|---|---|---|---|---|---|
| | | | | **1988** | | | | | | | | | | |
| Jaeger (Jaeger et al., 1988) | Literature Synthesis | 1985 | 2100 | *0.06° C per decade warming rate, -1.0 cm per decade SLR rate* | Low | | | | -1.15 | -1.0 | *Application of SLR rate* | | | |
| | | | | *0.3° C per decade warming rate, +5.5 cm per decade SLR rate* | Mid | | | | 0.6325 | 0.55 | | | | |
| | | | | *0.8° C per decade warming rate, +24 cm per decade SLR rate* | High | | | | 2.76 | 2.4 | | | | |
| van der Veen (van der Veen, 1988) | Literature Synthesis | 1985 | 2085 | *2.0°C warming in response to 2xCO2 that occurs in 2085* | Low | | | | 0.28 | 0.28 | *Sum of component estimates* | | | |
| | | | | *4.0°C warming in response to 2xCO2 that occurs in 2085* | High | | | | 0.66 | 0.66 | | | | |
| | | | | **1989** | | | | | | | | | | |
| Oerlemans (Oerlemans, 1989) | Model Synthesis | 1985 | 2100 | *After Jaeger (1988)* | High | | | | 0.656 | 0.57 | *Sum of "most likely" component contributions* | | | |
| | | | | **1990** | | | | | | | | | | |
| Meier (Meier, 1990) | Literature Synthesis | | 2050 | *3-5°C warming by 2050* | High | -0.08 | | *- 0.42 of central estimate* | 0.34 | | *Sum of component estimates* | 0.76 | | *+ 0.42 of central estimate* |
| Warrick (Warrick & Oerlemans, 1990) | IPCC Assessment Report | 1990 | 2100 | IPCC BAU | High | 0.31 | 0.28 | *1.5°C at 2xCO2* | 0.66 | 0.6 | *2.5°C at 2xCO2* | 1.1 | 1.0 | *4.5°C at 2xCO2* |
| | | | | IPCC B | Mid | 0.22 | 0.2 | | 0.47 | 0.43 | | 0.78 | 0.71 | |
| | | | | IPCC C | Mid | 0.18 | 0.16 | | 0.4 | 0.36 | | 0.67 | 0.61 | |
| | | | | IPCC D | Low | 0.16 | 0.15 | | 0.34 | 0.31 | | 0.59 | 0.54 | |
| | | | | **1991** | | | | | | | | | | |
| Budd (Budd & Simmonds, 1991) | Model Hybrid | 1990 | 2050 | *Doubling of CO2 by 2030* | High | | | | 0.27 | 0.45 | *Sum of component estimates* | | | |
| | | | | **1992** | | | | | | | | | | |
| Wigley (Wigley & Raper, 1992) | Model Synthesis | 1990 | 2100 | IS92a | Mid | | | | 0.48 | 0.44 | *"Best guess" model parameter values* | | | |
| | | | | IS92b | Mid | | | | 0.47 | 0.43 | | | | |
| | | | | IS92c | Low | | | | 0.35 | 0.32 | | | | |
| | | | | IS92d | Low | | | | 0.4 | 0.36 | | | | |
| | | | | IS92e | High | | | | 0.53 | 0.48 | | | | |
| | | | | IS92f | Mid | | | | 0.52 | 0.47 | | | | |
| | | | | **1993** | | | | | | | | | | |
| Wigley (Wigley & Raper, 1993) | Model Synthesis | 1990 | 2100 | IPCC BAU | High | 0.13 | 0.12 | *Model parameter values for extreme low T change* | 0.63 | 0.57 | *Model parameter values for mid-value T change* | 1.24 | 1.1 | *Model parameter values for extreme high T change* |
| | | | | IPCC B | Mid | 0.06 | 0.055 | | 0.46 | 0.42 | | 0.95 | 0.86 | |
| | | | | IPCC C | Mid | 0.03 | 0.027 | | 0.38 | 0.35 | | 0.84 | 0.76 | |

| Lead Author | Method | Base Year(s) | End Year(s) | Emission Scenario | Scen. Cat. | Lower Est. (m) | Lower Rate (m/cen) | Lower Estimate Definition | Central Est. (m) | Central Rate (m/cen) | Central Estimate Definition | Upper Est. (m) | Upper Rate (m/cen) | Upper Estimate Definition |
|---|---|---|---|---|---|---|---|---|---|---|---|---|---|---|
| **1995** | | | | | | | | | | | | | | |
| Titus (Titus & Narayanan, 1995) | Probabilistic | 1990 | 2100 | *Probability density functions based on IS92a-f scenarios updated by Wigley and Raper (1992)* | Mid | -0.286 | -0.26 | *2.5$^{th}$ percentile* | 0.337 | 0.31 | *Mean estimate* | 1.14 | 1.0 | *97.5$^{th}$ percentile* |
| Wigley (Wigley, 1995) | Model Synthesis | 1990 | 2100 | *CO2 concentration stabilization at 350 ppmv, 2.5°C at 2xCO2* | Low | | | | 0.15 | 0.14 | *Model output under specified scenario* | | | |
| | | | | *CO2 concentration stabilization at 450 ppmv, 2.5°C at 2xCO2* | Low | | | | 0.26 | 0.24 | | | | |
| | | | | *CO2 concentration stabilization at 550 ppmv, 2.5°C at 2xCO2* | Mid | | | | 0.32 | 0.29 | | | | |
| | | | | *CO2 concentration stabilization at 650 ppmv, 2.5°C at 2xCO2* | Mid | | | | 0.37 | 0.34 | | | | |
| | | | | *CO2 concentration stabilization at 750 ppmv, 2.5°C at 2xCO2* | Mid | | | | 0.41 | 0.37 | | | | |
| **1996** | | | | | | | | | | | | | | |
| Raper (Raper et al., 1996) | Model Synthesis | 1990 | 2100 | *IS95a, 1.5°C at 2xCO2, Low ice melt param. vals.* | Mid | | | | 0.2 | 0.18 | *Model output under specified scenario* | | | |
| | | | | *IS95a, 1.5°C at 2xCO2, Mid ice melt param. vals.* | Mid | | | | 0.35 | 0.32 | | | | |
| | | | | *IS95a, 2.5°C at 2xCO2, Low ice melt param. vals.* | Mid | | | | 0.198 | 0.18 | | | | |
| | | | | *IS95a, 2.5°C at 2xCO2, Mid ice melt param. vals.* | Mid | | | | 0.49 | 0.45 | | | | |
| | | | | *IS95a, 2.5°C at 2xCO2, High ice melt param. vals.* | Mid | | | | 0.86 | 0.78 | | | | |
| | | | | *IS95a, 4.5°C at 2xCO2, Mid ice melt param. vals.* | Mid | | | | 0.66 | 0.6 | | | | |

| Lead Author | Method | Base Year(s) | End Year(s) | Emission Scenario | Scen. Cat. | Lower Est. (m) | Lower Rate (m/cen) | Lower Estimate Definition | Central Est. (m) | Central Rate (m/cen) | Central Estimate Definition | Upper Est. (m) | Upper Rate (m/cen) | Upper Estimate Definition |
|---|---|---|---|---|---|---|---|---|---|---|---|---|---|---|
| Raper (Raper et al., 1996) | Model Synthesis | 1990 | 2100 | *IS95a, 4.5°C at 2xCO2, High ice melt param. vals.* | Mid | | | | 0.86 | 0.78 | *Model output under specified scenario* | | | |
| | | | | *IS95b, 2.5°C at 2xCO2, Mid ice melt param. vals.* | Mid | | | | 0.47 | 0.43 | | | | |
| | | | | *IS95b, 2.5°C at 2xCO2, Mid ice melt param. vals.* | Mid | | | | 0.38 | 0.35 | | | | |
| | | | | *IS95b, 2.5°C at 2xCO2, Mid ice melt param. vals.* | Mid | | | | 0.42 | 0.38 | | | | |
| | | | | *IS95b, 2.5°C at 2xCO2, Mid ice melt param. vals.* | Mid | | | | 0.56 | 0.51 | | | | |
| | | | | *IS95b, 2.5°C at 2xCO2, Mid ice melt param. vals.* | Mid | | | | 0.54 | 0.49 | | | | |
| | | | | *CO2 concentration stabilization at 450 ppmv, 1.5°C at 2xCO2, Low ice melt param. vals.* | Low | | | | 0.08 | 0.073 | | | | |
| | | | | *CO2 concentration stabilization at 450 ppmv, 2.5°C at 2xCO2, Mid ice melt param. vals.* | Mid | | | | 0.83 | 0.75 | | | | |
| | | | | *CO2 concentration stabilization at 450 ppmv, 4.5°C at 2xCO2, High ice melt param. vals.* | High | | | | 2.02 | 1.8 | | | | |
| | | | | *CO2 concentration stabilization at 650 ppmv, 1.5°C at 2xCO2, Low ice melt param. vals.* | Low | | | | 0.3 | 0.27 | | | | |
| | | | | *CO2 concentration stabilization at 650 ppmv, 2.5°C at 2xCO2, Mid ice melt param. vals.* | Mid | | | | 1.4 | 1.3 | | | | |
| | | | | *CO2 concentration stabilization at 650 ppmv, 4.5°C at 2xCO2, High ice melt param. vals.* | High | | | | 3.22 | 2.9 | | | | |
| Warrick (Warrick et al., 1996) | IPCC Assessment Report | 1990 | 2100 | *IS92a* | Mid | 0.2 | 0.18 | | 0.49 | 0.45 | | 0.86 | 0.78 | |

| Lead Author | Method | Base Year(s) | End Year(s) | Emission Scenario | Scen. Cat. | Lower Est. (m) | Lower Rate (m/cen) | Lower Estimate Definition | Central Est. (m) | Central Rate (m/cen) | Central Estimate Definition | Upper Est. (m) | Upper Rate (m/cen) | Upper Estimate Definition |
|---|---|---|---|---|---|---|---|---|---|---|---|---|---|---|
| | | | | **1997** | | | | | | | | | | |
| de Wolde (de Wolde et al., 1997) | Model Synthesis | 1990 | 2100 | *IS92a aerosols constant* | Mid | | | | 0.34 | 0.31 | | | | |
| | | | | *IS92b aerosols constant* | Mid | | | | 0.33 | 0.3 | | | | |
| | | | | *IS92c aerosols constant* | Low | | | | 0.23 | 0.21 | | | | |
| | | | | *IS92d aerosols constant* | Low | | | | 0.25 | 0.23 | | | | |
| | | | | *IS92e aerosols constant* | High | | | | 0.42 | 0.38 | | | | |
| | | | | *IS92f aerosols constant* | Mid | | | | 0.39 | 0.35 | | | | |
| | | | | *IS92a aerosols changing* | Mid | | | | 0.27 | 0.25 | | | | |
| | | | | *IS92b aerosols changing* | Mid | | | | 0.27 | 0.25 | | | | |
| | | | | *IS92c aerosols changing* | Low | | | | 0.23 | 0.21 | | | | |
| | | | | *IS92d aerosols changing* | Low | | | | 0.26 | 0.24 | | | | |
| | | | | *IS92e aerosols changing* | High | | | | 0.29 | 0.26 | | | | |
| | | | | *IS92f aerosols changing* | Mid | | | | 0.29 | 0.26 | | | | |
| | | | | **2000** | | | | | | | | | | |
| Gregory (Gregory & Lowe, 2000) | Model Synthesis | 1990 | 2100 | *IS92a* | Mid | | | | 0.48 | 0.44 | *Model output* | | | |
| | | | | *IS92a* | Mid | | | | 0.44 | 0.4 | | | | |
| | | | | **2001** | | | | | | | | | | |
| Church (Church et al., 2001) | IPCC Assessment Report | 1990 | 2090 | *SRES A1B* | Mid | 0.13 | 0.13 | | | | | 0.695 | 0.70 | |
| | | | | *SRES A1FI* | High | 0.18 | 0.18 | | | | | 0.86 | 0.86 | |
| | | | | *SRES A1T* | Mid | 0.115 | 0.12 | | | | | 0.67 | 0.67 | |
| | | | | *SRES A2* | High | 0.155 | 0.16 | | | | | 0.745 | 0.75 | |
| | | | | *SRES B1* | Mid | 0.09 | 0.09 | | | | | 0.57 | 0.57 | |
| | | | | *SRES B2* | Mid | 0.12 | 0.12 | | | | | 0.65 | 0.65 | |
| | | | | **2007** | | | | | | | | | | |
| Meehl (Meehl et al., 2007) | IPCC Assessment Report | 1980-1999 | 2090-2099 | *SRES B1* | Mid | 0.18 | 0.17 | 5[th] percentile of spread of model results | | | | 0.38 | 0.36 | 95[th] percentile of spread of model results |
| | | | | *SRES A1T* | Mid | 0.2 | 0.19 | | | | | 0.45 | 0.43 | |
| | | | | *SRES B2* | Mid | 0.2 | 0.19 | | | | | 0.43 | 0.41 | |
| | | | | *SRES A1B* | Mid | 0.21 | 0.2 | | | | | 0.48 | 0.46 | |
| | | | | *SRES A2* | High | 0.23 | 0.22 | | | | | 0.51 | 0.49 | |
| | | | | *SRES A1FI* | High | 0.26 | 0.25 | | | | | 0.59 | 0.56 | |
| | | | | *SRES B1\** | Mid | 0.18 | 0.17 | | | | | 0.47 | 0.44 | |
| | | | | *SRES A1T \** | Mid | 0.19 | 0.18 | | | | | 0.58 | 0.55 | |
| | | | | *SRES B2\** | Mid | 0.2 | 0.19 | | | | | 0.54 | 0.51 | |
| | | | | *SRES A1B\** | Mid | 0.2 | 0.19 | | | | | 0.61 | 0.58 | |
| | | | | *SRES A2\** | High | 0.22 | 0.21 | | | | | 0.64 | 0.61 | |
| | | | | *SRES A1FI\** | High | 0.25 | 0.24 | | | | | 0.76 | 0.72 | |
| Rahmstorf (Rahmstorf, 2007) | Semi-empirical | 1990 | 2100 | *SRES B1* | Mid | | | | 0.5 | 0.45 | | | | |
| | | | | *SRES A1FI* | High | | | | 1.4 | 1.3 | | | | |

---

\* With scaled up ice sheet discharge

| Lead Author | Method | Base Year(s) | End Year(s) | Emission Scenario | Scen. Cat. | Lower Est. (m) | Lower Rate (m/cen) | Lower Estimate Definition | Central Est. (m) | Central Rate (m/cen) | Central Estimate Definition | Upper Est. (m) | Upper Rate (m/cen) | Upper Estimate Definition |
|---|---|---|---|---|---|---|---|---|---|---|---|---|---|---|
| | | | | | | | | **2008** | | | | | | |
| Cayan (Cayan et al., 2008) | Model Synthesis | 1990 | 2070-2099 | SRES A1FI | High | 0.168 | 0.18 | *Low estimate of land-based ice from MAGICC* | | | | 0.716 | 0.76 | *High estimate of land-based ice from MAGICC* |
| | | | | SRES A2 | High | 0.142 | 0.15 | | | | | 0.605 | 0.64 | |
| | | | | SRES B1 | Mid | 0.109 | 0.12 | | | | | 0.539 | 0.57 | |
| Horton, R. (R. Horton et al., 2008) | Semi-empirical | 2001-2005 | 2100 | SRES A2 | High | 0.68 | 0.70 | *Minimum of model spread* | | | | 0.89 | 0.92 | *Maximum of model spread* |
| | | | | SRES A1B | Mid | 0.62 | 0.64 | | | | | 0.88 | 0.91 | |
| | | | | SRES B1 | Mid | 0.54 | 0.56 | | | | | 0.75 | 0.77 | |
| Pfeffer (Pfeffer et al., 2008) | Model Hybrid | 2007 | 2100 | Low 1 | Low | | | | 0.785 | 0.84 | *Sum of component estimates* | | | |
| | | | | Low 2 | Low | | | | 0.833 | 0.90 | | | | |
| | | | | High 1 | High | | | | 2.008 | 2.2 | | | | |
| | | | | | | | | **2009** | | | | | | |
| Vermeer (Vermeer & Rahmstorf, 2009) | Semi-empirical | 1990 | 2100 | SRES B1 | Mid | 0.81 | 0.74 | *- 7% (1 std dev)* | | | | 1.31 | 1.19 | *+ 7% (1 std dev)* |
| | | | | SRES A1T | Mid | 0.97 | 0.88 | | | | | 1.58 | 1.44 | |
| | | | | SRES B2 | Mid | 0.89 | 0.81 | | | | | 1.45 | 1.32 | |
| | | | | SRES A1B | Mid | 0.97 | 0.88 | | | | | 1.56 | 1.42 | |
| | | | | SRES A2 | High | 0.98 | 0.89 | | | | | 1.55 | 1.41 | |
| | | | | SRES A1FI | High | 1.13 | 1.0 | | | | | 1.79 | 1.63 | |
| | | | | | | | | **2010** | | | | | | |
| Grinsted (Grinsted et al., 2010) | Semi-empirical | 1980-1999 | 2090-2099 | SRES A1B[†] | Mid | 0.91 | 0.87 | *5th percentile* | | | | 1.32 | 1.26 | *95th percentile* |
| | | | | SRES A1FI[†] | High | 1.1 | 1.05 | | | | | 1.6 | 1.52 | |
| | | | | SRES A1T[†] | Mid | 0.89 | 0.85 | | | | | 1.3 | 1.24 | |
| | | | | SRES A2[†] | High | 0.93 | 0.89 | | | | | 1.36 | 1.30 | |
| | | | | SRES B1[†] | Mid | 0.72 | 0.69 | | | | | 1.07 | 1.02 | |
| | | | | SRES B2[†] | Mid | 0.82 | 0.78 | | | | | 1.2 | 1.14 | |
| | | | | T constant at 1980-1999 average[†] | Low | 0.21 | 0.2 | | | | | 0.38 | 0.36 | |
| | | | | SRES A1B[‡] | Mid | 1.21 | 1.15 | | | | | 1.79 | 1.70 | |
| | | | | SRES A1FI[‡] | High | 1.45 | 1.38 | | | | | 2.15 | 2.05 | |
| | | | | SRES A1T[‡] | Mid | 1.18 | 1.12 | | | | | 1.76 | 1.68 | |
| | | | | SRES A2[‡] | High | 1.24 | 1.18 | | | | | 1.83 | 1.74 | |
| | | | | SRES B1[‡] | Mid | 0.96 | 0.91 | | | | | 1.44 | 1.37 | |
| | | | | SRES B2[‡] | Mid | 1.09 | 1.04 | | | | | 1.62 | 1.54 | |
| | | | | T constant at 1980-1999 average[‡] | Low | 0.29 | 0.28 | | | | | 0.49 | 0.47 | |
| | | | | SRES A1B[§] | Mid | 0.32 | 0.30 | | | | | 1.34 | 1.28 | |
| | | | | SRES A1FI[§] | High | 0.34 | 0.32 | | | | | 1.59 | 1.51 | |
| | | | | SRES A1T[§] | Mid | 0.32 | 0.30 | | | | | 1.32 | 1.26 | |
| | | | | SRES A2[§] | High | 0.32 | 0.30 | | | | | 1.37 | 1.30 | |
| | | | | SRES B1[§] | Mid | 0.3 | 0.29 | | | | | 1.1 | 1.05 | |
| | | | | SRES B2[§] | Mid | 0.31 | 0.30 | | | | | 1.22 | 1.16 | |
| | | | | T constant at 1980-1999 average[§] | Low | 0.22 | 0.21 | | | | | 0.44 | 0.42 | |

---

[†] Using 'Moberg' model parameters
[‡] Using 'Jones and Mann' model parameters
[§] Using 'Historical only' model parameters

| Lead Author | Method | Base Year(s) | End Year(s) | Emission Scenario | Scen. Cat. | Lower Est. (m) | Lower Rate (m/cen) | Lower Estimate Definition | Central Est. (m) | Central Rate (m/cen) | Central Estimate Definition | Upper Est. (m) | Upper Rate (m/cen) | Upper Estimate Definition |
|---|---|---|---|---|---|---|---|---|---|---|---|---|---|---|
| Hunter (Hunter, 2010) | Semi-empirical | 1990 | 2100 | SRES A1B | Mid | 0.208 | 0.19 | *5th percentile* | | | | 0.649 | 0.59 | *95th percentile* |
| | | | | SRES A1T | Mid | 0.194 | 0.18 | | | | | 0.611 | 0.56 | |
| | | | | SRES A1FI | High | 0.266 | 0.24 | | | | | 0.819 | 0.74 | |
| | | | | SRES A2 | High | 0.237 | 0.26 | | | | | 0.692 | 0.63 | |
| | | | | SRES B1 | Mid | 0.185 | 0.17 | | | | | 0.496 | 0.45 | |
| | | | | SRES B2 | Mid | 0.21 | 0.19 | | | | | 0.576 | 0.52 | |
| Jevrejeva (Jevrejeva et al., 2010) | Semi-empirical | 1980-2000 | 2100 | SRES A1B | Mid | 0.2 | 0.18 | *5th percentile* | | | | 1.4 | 1.27 | *95th percentile* |
| | | | | SRES A1FI | High | 0.8 | 0.73 | | | | | 1.6 | 1.45 | |
| | | | | SRES A1T | Mid | 0.7 | 0.64 | | | | | 1.3 | 1.18 | |
| | | | | SRES A2 | High | 1.3 | 1.18 | | | | | 1.5 | 1.36 | |
| | | | | SRES B1 | Mid | 0.6 | 0.55 | | | | | 1.1 | 1.00 | |
| | | | | SRES B2 | Mid | 0.7 | 0.64 | | | | | 1.3 | 1.18 | |
| Moore (Moore et al., 2010) | Semi-empirical | 1980-2000 | 2100 | RCP2.6 | Low | 0.37 | 0.34 | *5th percentile* | 0.55 | 0.5 | *Median* | 0.8 | 0.73 | *95th percentile* |
| | | | | RCP4.5 | Mid | 0.59 | 0.54 | | 0.85 | 0.77 | | 1.2 | 1.09 | |
| | | | | RCP8.5 | High | 0.8 | 0.73 | | 1.1 | 1.0 | | 1.5 | 1.36 | |
| | | | | SRES A1B | Mid | 0.65 | 0.59 | | 0.95 | 0.86 | | 1.35 | 1.23 | |
| | | | | SRES A1FI | High | 0.85 | 0.77 | | 1.15 | 1.05 | | 1.65 | 1.5 | |
| | | | | SRES A1T | Mid | 0.6 | 0.55 | | 0.85 | 0.77 | | 1.25 | 1.14 | |
| | | | | SRES A2 | High | 0.7 | 0.64 | | 1 | 0.91 | | 1.45 | 1.32 | |
| | | | | SRES B1 | Mid | 0.5 | 0.45 | | 0.7 | 0.64 | | 1.05 | 0.95 | |
| | | | | SRES B2 | Mid | 0.58 | 0.53 | | 0.85 | 0.77 | | 1.2 | 1.09 | |
| **2011** | | | | | | | | | | | | | | |
| Katsman (Katsman et al., 2011) | Model Hybrid | 1990 | 2100 | | None | 0.55 | 0.5 | | | | | 1.15 | 1.05 | |
| Pardaens (Pardaens et al., 2011) | Model Synthesis | 1980-1999 | 2090-2099 | SRES A1B | Mid | 0.23 | 0.22 | *5th percentile* | | | | 0.43 | 0.41 | *95th percentile* |
| | | | | SRES A1B(IMAGE) | Mid | 0.29 | 0.28 | | | | | 0.51 | 0.49 | |
| | | | | E1 | Low | 0.17 | 0.16 | | | | | 0.34 | 0.32 | |
| USACE (U.S. Army Corps of Engineers, 2011) | Literature Synthesis | 1992 | 2100 | | None | 0.5 | 0.46 | | | | | 1.5 | 1.38 | |
| **2012** | | | | | | | | | | | | | | |
| Jevrejeva (Jevrejeva et al., 2012) | Semi-empirical | 1980-2000 | 2100 | RCP8.5 | High | 0.81 | 0.74 | *5th percentile* | | | | 1.65 | 1.50 | *95th percentile* |
| | | | | RCP6.0 | Mid | 0.6 | 0.55 | | | | | 1.26 | 1.15 | |
| | | | | RCP4.5 | Mid | 0.52 | 0.47 | | | | | 1.1 | 1.0 | |
| | | | | RCP2.6 | Low | 0.36 | 0.33 | | | | | 0.83 | 0.75 | |
| National Research Council (National Research Council, 2012) | Model Hybrid | 2000 | 2030 | SRES B1 - A1FI | Mid | 0.083 | 0.28 | *Subtracting twice the standard deviation from the mean projection, and adjusting to the difference between A1B and B1* | 0.135 | 0.45 | *Mean estimate* | 0.232 | 0.77 | *Adding twice the standard deviation to the mean, adjusting to the difference between A1FI and A1B, and adding the dynamical imbalance contribution* |
| | | | 2050 | SRES B1 - A1FI | Mid | 0.176 | 0.35 | | 0.28 | 0.56 | *Mean estimate* | 0.482 | 0.96 | |
| | | | 2100 | SRES B1 - A1FI | Mid | 0.504 | 0.50 | | 0.827 | 0.83 | *Mean estimate* | 1.402 | 1.40 | |

| Lead Author | Method | Base Year(s) | End Year(s) | Emission Scenario | Scen. Cat. | Lower Est. (m) | Lower Rate (m/cen) | Lower Estimate Definition | Central Est. (m) | Central Rate (m/cen) | Central Estimate Definition | Upper Est. (m) | Upper Rate (m/cen) | Upper Estimate Definition |
|---|---|---|---|---|---|---|---|---|---|---|---|---|---|---|
| Parris (Parris et al., 2012) | Literature Synthesis | 1992 | 2100 | *Highest--estimated ocean warming from AR4 and max glacier and ice sheet loss by 2100* | None | | | | 2 | 1.85 | *Based on lit review* | | | |
| | | | | *Average of High-end of Semi-empirical, global SLR projections* | None | | | | 1.2 | 1.11 | | | | |
| | | | | *Upper end of AR4 projections from climate models using B1 scenario* | None | | | | 0.5 | 0.46 | | | | |
| | | | | *Linear extrapolation of the historical SLR rate* | None | | | | 0.2 | 0.19 | | | | |
| PICRCA (World Bank, 2012) | Model Synthesis | 1980-1999 | 2090-2099 | 2°C Lower ice sheet | Low | 0.27 | 0.26 | 16th percentile | 0.34 | 0.32 | *Sum of component estimates* | 0.42 | 0.4 | 84th percentile |
| | | | | 2°C Semi-empirical | Low | 0.65 | 0.62 | | 0.79 | 0.75 | | 0.96 | 0.91 | |
| | | | | 4°C Lower ice sheet | High | 0.37 | 0.35 | | 0.47 | 0.45 | | 0.58 | 0.55 | |
| | | | | 4°C Semi-empirical | High | 0.82 | 0.78 | | 0.96 | 0.91 | | 1.23 | 1.17 | |
| Rahmstorf (Rahmstorf et al., 2012) | Semi-empirical | 2000 | 2100 | RCP4.5 | Mid | 0.45 | 0.45 | 5[th] percentile | | | | 1.39 | 1.39 | 95[th] percentile |
| | | | | RCP8.5 | High | 0.55 | 0.55 | | | | | 2.03 | 2.03 | |
| Schaeffer (Schaeffer et al., 2012) | Semi-empirical | 2000 | 2100 | CPH reference | Mid | 0.72 | 0.72 | 5[th] percentile | 1.02 | 1.02 | *Median* | 1.39 | 1.39 | 95[th] percentile |
| | | | | CPH policy | Mid | 0.68 | 0.68 | | 0.96 | 0.96 | | 1.32 | 1.32 | |
| | | | | RCP4.5 | Mid | 0.64 | 0.64 | | 0.9 | 0.9 | | 1.21 | 1.21 | |
| | | | | Stab 2°C | Low | 0.56 | 0.56 | | 0.8 | 0.8 | | 1.05 | 1.05 | |
| | | | | RCP2.6 | Low | 0.52 | 0.52 | | 0.75 | 0.75 | | 0.96 | 0.96 | |
| | | | | MERGE400 | Low | 0.54 | 0.54 | | 0.77 | 0.77 | | 0.99 | 0.99 | |
| | | | | Zero 2016 | Low | 0.4 | 0.4 | | 0.59 | 0.59 | | 0.8 | 0.8 | |
| Sriver (Sriver et al., 2012) | Model Hybrid | 1950-2003 | 2100 | RCP8.5 | High | | | | 2.25 | 1.82 | | | | |
| Zecca (Zecca & Chiari, 2012) | Semi-empirical | 2000 | 2100 | High | High | 0.73 | 0.73 | | 0.94 | 0.94 | *Mean* | 1.2 | 1.2 | |
| | | | | medium | Mid | 0.67 | 0.67 | | 0.84 | 0.84 | | 1.07 | 1.07 | |
| | | | | Low | Low | 0.63 | 0.63 | | 0.78 | 0.78 | | 0.99 | 0.99 | |
| | | | | mitigation | Mid | 0.56 | 0.56 | | 0.66 | 0.66 | | 0.82 | 0.82 | |
| **2013** | | | | | | | | | | | | | | |
| Bamber (Bamber & Aspinall, 2013) | Expert Judgement | 2010 | 2100 | | None | 0.33 | 0.37 | 5[th] percentile | | | | 1.32 | 1.47 | 95[th] percentile |
| Church (Church et al., 2013) | IPCC Assessment Report | 1986-2005 | 2100 | RCP2.6 | Low | 0.28 | 0.27 | 17th percentile | 0.44 | 0.42 | *Median* | 0.61 | 0.58 | 83rd percentile |
| | | | | RCP4.5 | Mid | 0.36 | 0.34 | | 0.53 | 0.51 | | 0.71 | 0.68 | |
| | | | | RCP6.0 | Mid | 0.38 | 0.36 | | 0.55 | 0.53 | | 0.73 | 0.70 | |
| | | | | RCP8.5 | High | 0.52 | 0.50 | | 0.74 | 0.70 | | 0.98 | 0.94 | |
| Houston (Houston, 2013) | Other | 1990 | 2100 | | None | 0.18 | 0.16 | | | | | 0.82 | 0.75 | |

| Lead Author | Method | Base Year(s) | End Year(s) | Emission Scenario | Scen. Cat. | Lower Est. (m) | Lower Rate (m/cen) | Lower Estimate Definition | Central Est. (m) | Central Rate (m/cen) | Central Estimate Definition | Upper Est. (m) | Upper Rate (m/cen) | Upper Estimate Definition |
|---|---|---|---|---|---|---|---|---|---|---|---|---|---|---|
| Miller (Miller et al., 2013) | Model Hybrid | 2000 | 2100 | *Central* | Mid | | | | 0.77 | 0.77 | | | | |
| | | | | *Low* | Mid | | | | 0.43 | 0.43 | | | | |
| | | | | *High* | High | | | | 1.21 | 1.21 | | | | |
| | | | | *Higher* | High | | | | 1.41 | 1.41 | | | | |
| Orlić (Orlić & Pasarić, 2013) | Semi-empirical | 2000 | 2100 | *SRES B1*[**] | Mid | 0.68 | 0.68 | *Standard deviation* | 0.74 | 0.74 | *Mean estimate* | 0.8 | 0.8 | *Standard deviation* |
| | | | | *SRES B1*[††] | Mid | 0.9 | 0.9 | | 0.95 | 0.95 | | 1 | 1 | |
| | | | | *SRES B1*[‡‡] | Mid | 0.55 | 0.55 | | 0.62 | 0.62 | | 0.69 | 0.69 | |
| | | | | *SRES B1*[§§] | Mid | 0.59 | 0.59 | | 0.74 | 0.74 | | 0.89 | 0.89 | |
| | | | | *SRES B1*[***] | Mid | 0.78 | 0.78 | | 0.95 | 0.95 | | 1.12 | 1.12 | |
| | | | | *SRES B1*[†††] | Mid | 0.48 | 0.48 | | 0.62 | 0.62 | | 0.76 | 0.76 | |
| Perrette (Perrette et al., 2013) | Semi-empirical | 1980-1999 | 2080-2099 | *RCP2.6* | Low | 0.59 | 0.59 | *16th percentile* | 0.75 | 0.75 | *Mean estimate* | 0.94 | 0.94 | *84th percentile* |
| | | | | *RCP4.5* | Mid | 0.66 | 0.66 | | 0.86 | 0.86 | | 1.11 | 1.11 | |
| | | | | *RCP6.0.0* | Mid | 0.66 | 0.66 | | 0.86 | 0.86 | | 1.1 | 1.1 | |
| | | | | *RCP8.5* | High | 0.78 | 0.78 | | 1.06 | 1.06 | | 1.43 | 1.43 | |
| **2014** | | | | | | | | | | | | | | |
| Horton (B. P. Horton et al., 2014) | Expert Judgement | 1986-2005 | 2100 | *RCP2.6* | Low | 0.25 | 0.24 | *5th percentile* | | | | 0.7 | 0.67 | *95th percentile* |
| | | | | *RCP8.5* | High | 0.5 | 0.48 | | | | | 1.5 | 1.44 | |
| Jevrejeva (Jevrejeva et al., 2014) | Probabilistic | 2000 | 2100 | *RCP8.5* | High | 0.46 | 0.46 | *5th percentile* | 0.8 | 0.8 | *Median* | 1.8 | 1.8 | *95th percentile* |
| Kopp (Kopp et al., 2014) | Probabilistic | 2000 | 2100 | *RCP2.6* | Low | 0.29 | 0.29 | *5th percentile* | 0.5 | 0.5 | *Median* | 0.82 | 0.82 | *95th percentile* |
| | | | | *RCP4.5* | Mid | 0.36 | 0.36 | | 0.59 | 0.59 | | 0.93 | 0.93 | |
| | | | | *RCP8.5* | High | 0.52 | 0.52 | | 0.79 | 0.79 | | 1.21 | 1.21 | |
| Slangen (Slangen et al., 2014) | Model Synthesis | 1986-2005 | 2081-2100 | *RCP4.5* | Mid | 0.35 | 0.37 | *16th percentile* | 0.54 | 0.57 | *Median* | 0.73 | 0.77 | *84th percentile* |
| | | | | *RCP8.5* | High | 0.43 | 0.45 | | 0.71 | 0.75 | | 0.99 | 1.04 | |
| **2015** | | | | | | | | | | | | | | |
| Grinsted (Grinsted et al., 2015) | Probabilistic | 2000 | 2100 | *RCP8.5* | High | 0.45 | 0.45 | *5th percentile* | 0.8 | 0.8 | *Median* | 1.83 | 1.83 | *95th percentile* |
| **2016** | | | | | | | | | | | | | | |
| Jackson (Jackson & Jevrejeva, 2016) | Probabilistic | 1986-2005 | 2100 | *RCP4.5* | Mid | 0.22 | 0.21 | *5th percentile* | 0.54 | 0.52 | *Median* | 0.85 | 0.81 | *95th percentile* |
| | | | | *RCP8.5* | High | 0.37 | 0.35 | | 0.75 | 0.72 | | 1.18 | 1.13 | |
| | | | | *High end* | High | 0.51 | 0.49 | | 0.84 | 0.80 | | 1.67 | 1.60 | |
| Jevrejeva (Jevrejeva et al., 2016) | Probabilistic | 2005 | 2100 | *RCP4.5* | Mid | 0.4 | 0.42 | *5th percentile* | 0.5 | 0.53 | *Median* | 0.7 | 0.74 | *95th percentile* |
| | | | | *RCP8.5* | High | 0.53 | 0.56 | | 0.86 | 0.91 | | 1.78 | 1.87 | |
| Kopp (Kopp et al., 2016) | Semi-empirical | 2000 | 2100 | *RCP2.6* | Low | 0.24 | 0.24 | *5th percentile* | 0.38 | 0.38 | *Median* | 0.61 | 0.61 | *95th percentile* |
| | | | | *RCP4.5* | Mid | 0.33 | 0.33 | | 0.51 | 0.51 | | 0.85 | 0.85 | |
| | | | | *RCP8.5* | High | 0.52 | 0.52 | | 0.76 | 0.76 | | 1.31 | 1.31 | |

---

[**] No correction to coefficients

[††] Chao correction to coefficients

[‡‡] Pokhrel correction to coefficients

[§§] No correction to coefficients, temperature projection variability taken into account

[***] Chao correction to coefficients, temperature projection variability taken into account

[†††] Pokhrel correction to coefficients, temperature projection variability taken into account

| Lead Author | Method | Base Year(s) | End Year(s) | Emission Scenario | Scen. Cat. | Lower Est. (m) | Lower Rate (m/cen) | Lower Estimate Definition | Central Est. (m) | Central Rate (m/cen) | Central Estimate Definition | Upper Est. (m) | Upper Rate (m/cen) | Upper Estimate Definition |
|---|---|---|---|---|---|---|---|---|---|---|---|---|---|---|
| Mengel (Mengel et al., 2016) | Model Synthesis | 1986-2005 | 2100 | RCP2.6 | Low | 0.28 | 0.27 | 5th percentile | 0.39 | 0.37 | Median | 0.56 | 0.54 | 95th percentile |
| | | | | RCP8.5 | High | 0.57 | 0.55 | | 0.85 | 0.81 | | 1.31 | 1.25 | |
| | | | | RCP4.5 | Mid | 0.37 | 0.35 | | 0.53 | 0.51 | | 0.77 | 0.74 | |
| Schleussner (Schleussner et al., 2016) | Semi-empirical | 2000 | 2100 | 1.5 degrees C above pre-industrial warming | Low | 0.29 | 0.29 | 16th percentile | 0.41 | 0.41 | Median | 0.53 | 0.53 | 84th percentile |
| | | | | 2.0 degrees C above pre-industrial warming | Low | 0.36 | 0.36 | | 0.5 | 0.5 | | 0.65 | 0.65 | |
| **2017** | | | | | | | | | | | | | | |
| Bakker (Bakker et al., 2017) | Probabilistic | 1986-2005 | 2100 | RCP2.6 | Low | 0.4 | 0.38 | 5th percentile | 0.53 | 0.51 | Mean estimate | 0.71 | 0.68 | 95th percentile |
| | | | | RCP4.5 | Mid | 0.54 | 0.52 | | 0.715 | 0.68 | | 0.97 | 0.93 | |
| | | | | RCP8.5 | High | 0.85 | 0.81 | | 1.16 | 1.11 | | 1.59 | 1.52 | |
| | | | | RCP8.5‡‡‡ | High | 1.62 | 1.55 | | 1.95 | 1.87 | | 2.37 | 2.27 | |
| Bittermann (Bittermann et al., 2017) | Semi-empirical | 2000 | 2100 | Temps Stabilize at 1.5 deg. Above pre-industrial levels | Low | 0.29 | 0.29 | 5th percentile | 0.37 | 0.37 | Median | 0.46 | 0.46 | 95th percentile |
| | | | | Temps Stabilize at 2.0 deg. Above pre-industrial levels | Low | 0.39 | 0.39 | | 0.5 | 0.5 | | 0.61 | 0.61 | |
| | | | | Temps overshoot 2.0 deg. Above pre-industrial levels before returning to 1.5 deg above pre-industrial | Low | 0.34 | 0.34 | | 0.44 | 0.44 | | 0.55 | 0.55 | |
| de Winter (De Winter et al., 2017) | Probabilistic | 1990-2010 | 2100 | RCP8.5§§§ | High | | | | 0.95 | 0.95 | Median | | | |
| | | | | RCP8.5 | High | | | | 0.76 | 0.76 | | | | |
| Goodwin (Goodwin et al., 2017) | Probabilistic | 1986-2005 | 2100 | RCP2.6**** | Low | 0.31 | 0.30 | 5th percentile | 0.44 | 0.42 | Mean estimate | 0.59 | 0.56 | 95th percentile |
| | | | | RCP4.5**** | Mid | 0.38 | 0.36 | | 0.53 | 0.51 | | 0.7 | 0.67 | |
| | | | | RCP6.0**** | Mid | 0.41 | 0.39 | | 0.57 | 0.55 | | 0.74 | 0.71 | |
| | | | | RCP8.5**** | High | 0.53 | 0.51 | | 0.72 | 0.69 | | 0.91 | 0.87 | |
| | | | | RCP2.6†††† | Low | 0.45 | 0.43 | | 0.57 | 0.55 | | 0.7 | 0.67 | |
| | | | | RCP4.5†††† | Mid | 0.55 | 0.53 | | 0.68 | 0.65 | | 0.82 | 0.78 | |
| | | | | RCP6.0†††† | Mid | 0.59 | 0.56 | | 0.72 | 0.69 | | 0.86 | 0.82 | |
| | | | | RCP8.5†††† | High | 0.76 | 0.73 | | 0.89 | 0.85 | | 1.03 | 0.99 | |
| | | | | RCP2.6‡‡‡‡ | Low | 0.45 | 0.43 | | 0.57 | 0.55 | | 0.72 | 0.69 | |
| | | | | RCP4.5‡‡‡‡ | Mid | 0.55 | 0.53 | | 0.69 | 0.66 | | 0.84 | 0.80 | |
| | | | | RCP6.0‡‡‡‡ | Mid | 0.59 | 0.56 | | 0.73 | 0.70 | | 0.88 | 0.84 | |
| | | | | RCP8.5‡‡‡‡ | High | 0.76 | 0.73 | | 0.9 | 0.86 | | 1.05 | 1.00 | |

---

‡‡‡ Using DeConto and Pollard ice sheet contribution

§§§ Asymmetric melt of Antarctic Ice Sheet

**** Constraints derived from the historical performance of the CMIP5 ensemble

†††† Constraints derived from a combination of CMIP5 simulated histories and historic observations

‡‡‡‡ Constraints derived entirely from historic observations

| Lead Author | Method | Base Year(s) | End Year(s) | Emission Scenario | Scen. Cat. | Lower Est. (m) | Lower Rate (m/cen) | Lower Estimate Definition | Central Est. (m) | Central Rate (m/cen) | Central Estimate Definition | Upper Est. (m) | Upper Rate (m/cen) | Upper Estimate Definition |
|---|---|---|---|---|---|---|---|---|---|---|---|---|---|---|
| Kopp (Kopp et al., 2017) | Probabilistic | 2000 | 2100 | *RCP8.5* | High | 0.93 | 0.93 | *5th percentile* | 1.46 | 1.46 | *Median* | 2.43 | 2.43 | *95th percentile* |
| | | | | *RCP4.5* | Mid | 0.5 | 0.5 | | 0.91 | 0.91 | | 1.58 | 1.58 | |
| | | | | *RCP2.6* | Low | 0.26 | 0.26 | | 0.56 | 0.56 | | 0.98 | 0.98 | |
| Le Bars (Le Bars et al., 2017) | Probabilistic | 1986-2005 | 2100 | *RCP8.5* | High | 0.51 | 0.49 | *5th percentile* | 0.73 | 0.70 | *Median* | 0.98 | 0.94 | *95th percentile* |
| | | | | *RCP8.5* | High | 1.21 | 1.16 | | 1.84 | 1.76 | | 2.47 | 2.36 | |
| | | | | *RCP8.5‡‡‡* | High | 1.04 | 1.00 | | 1.84 | 1.76 | | 2.65 | 2.54 | |
| Nauels (Nauels, Meinshausen, et al., 2017) | Model Synthesis | 1986-2005 | 2081-2100 | *RCP2.6* | Low | 0.32 | 0.34 | *17th percentile* | 0.41 | 0.43 | *Median* | 0.51 | 0.54 | *83rd percentile* |
| | | | | *RCP4.5* | Mid | 0.41 | 0.43 | | 0.49 | 0.52 | | 0.6 | 0.63 | |
| | | | | *RCP6.0* | Mid | 0.4 | 0.42 | | 0.49 | 0.52 | | 0.62 | 0.65 | |
| | | | | *RCP8.5* | High | 0.55 | 0.58 | | 0.67 | 0.0071 | | 0.83 | 0.87 | |
| Nauels (Nauels, Rogelj, et al., 2017) | Probabilistic | 1986-2005 | 2081-2100 | *RCP2.6* | Low | 0.312 | 0.33 | *17th percentile* | 0.496 | 0.52 | *Median* | 0.83 | 0.87 | *83rd percentile* |
| | | | | *RCP4.5* | Mid | 0.405 | 0.43 | | 0.636 | 0.67 | | 0.944 | 0.99 | |
| | | | | *RCP6.0* | Mid | 0.495 | 0.52 | | 0.758 | 0.80 | | 1.078 | 1.13 | |
| | | | | *RCP8.5* | High | 0.562 | 0.59 | | 0.84 | 0.88 | | 1.209 | 1.27 | |
| Sweet (Sweet et al., 2017) | Literature Synthesis | 2000 | 2100 | *Low* | None | | | | 0.3 | 0.3 | *Median* | | | |
| | | | | *Intermediate Low* | None | | | | 0.5 | 0.5 | | | | |
| | | | | *Intermediate* | None | | | | 1 | 1 | | | | |
| | | | | *Intermediate High* | None | | | | 1.5 | 1.5 | | | | |
| | | | | *High* | None | | | | 2 | 2 | | | | |
| | | | | *Extreme* | None | | | | 2.5 | 2.5 | | | | |
| Wigley (Wigley, 2017) | Model Synthesis | 2000 | 2100 | *250 ppm stabilization* | Low | | | | 0.36 | 0.36 | *Mean estimate* | | | |
| | | | | *350 ppm stabilization* | Low | | | | 0.425 | 0.425 | | | | |
| | | | | *450 ppm stabilization* | Low | | | | 0.48 | 0.48 | | | | |
| Wong (Wong et al., 2017) | Probabilistic | 2000 | 2100 | *RCP2.6* | Low | 0.43 | 0.43 | *5th percentile* | 0.55 | 0.55 | *Median* | 0.74 | 0.74 | *95th percentile* |
| | | | | *RCP4.5* | Mid | 0.56 | 0.56 | | 0.77 | 0.77 | | 1.3 | 1.3 | |
| | | | | *RCP8.5* | High | 1.09 | 1.09 | | 1.5 | 1.5 | | 2.07 | 2.07 | |
| **2018** | | | | | | | | | | | | | | |
| Jackson (Jackson et al., 2018) | Probabilistic | 1986-2005 | 2100 | *T 1.5 °C above pre-industrial levels* | Low | 0.2 | 0.19 | *5th percentile* | 0.44 | 0.42 | *Median* | 0.67 | 0.64 | *95th percentile* |
| | | | | *T 2.0 °C above pre-industrial levels* | Low | 0.24 | 0.22 | | 0.5 | 0.48 | | 0.74 | 0.71 | |
| | | | | *T 1.5 °C above pre-industrial levels* | Low | 0.28 | 0.27 | | 0.57 | 0.55 | | 0.93 | 0.89 | |
| | | | | *T 2.0 °C above pre-industrial levels* | Low | 0.32 | 0.31 | | 0.68 | 0.65 | | 1.17 | 1.12 | |
| Nerem (Nerem et al., 2018) | Other | 2005 | 2100 | | None | 0.535 | 0.56 | *central estimate - 1sigma* | 0.654 | 0.69 | *central estimate* | 0.773 | 0.81 | *central estimate + 1sigma* |
| Rasmussen (Rasmussen et al., 2018) | Probabilistic | 2000 | 2100 | *Temps 1.5 degrees above pre-industrial levels* | Low | 0.28 | 0.28 | *5th percentile* | 0.47 | 0.47 | *Median* | 0.82 | 0.82 | *95th percentile* |
| | | | | *Temps 2.0 degrees above pre-industrial levels* | Low | 0.3 | 0.3 | | 0.55 | 0.55 | | 0.94 | 0.94 | |
| | | | | *Temps 2.5 degrees above pre-industrial levels* | Mid | 0.36 | 0.36 | | 0.58 | 0.58 | | 0.93 | 0.93 | |

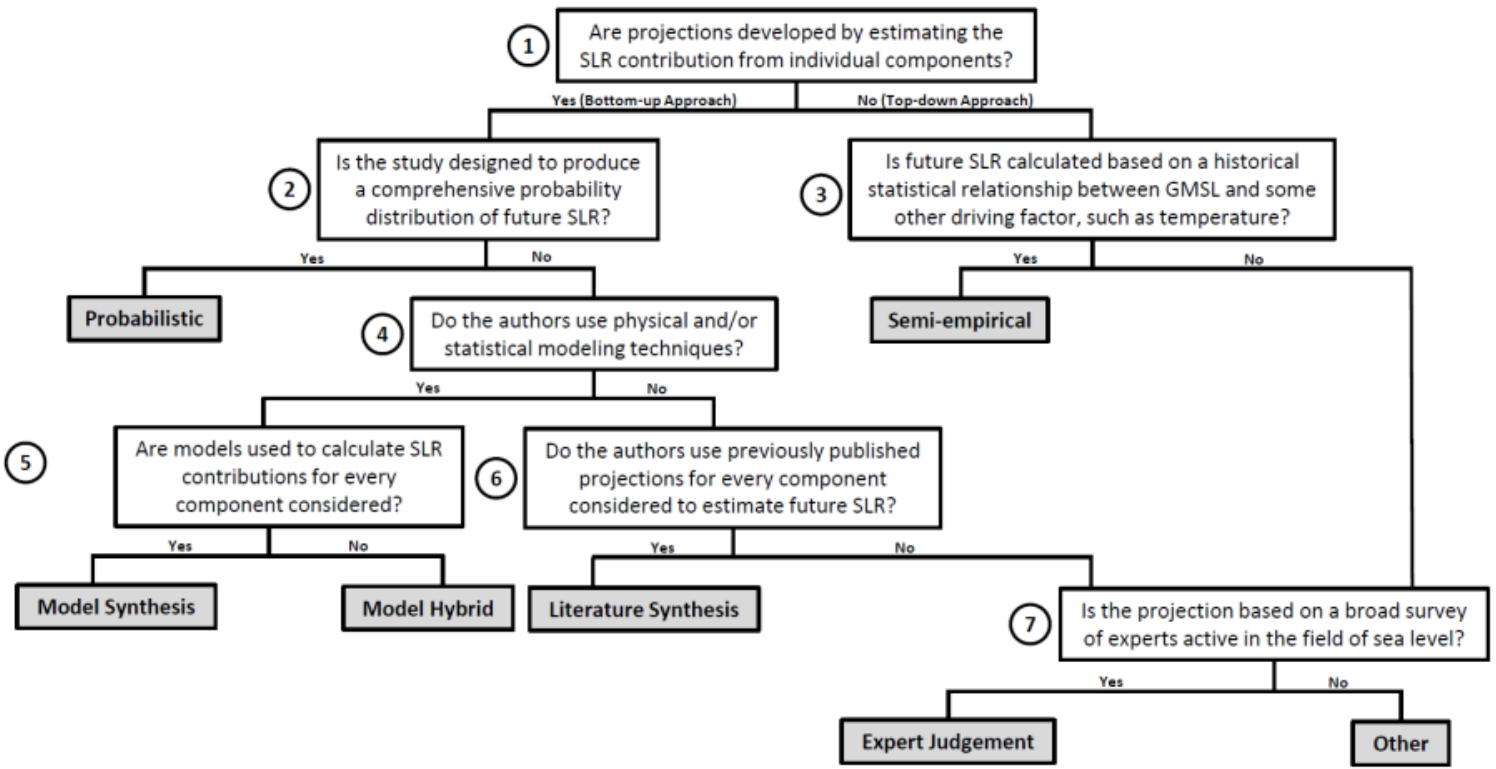

**Figure S1:** Decision tree showing the decision rules used to classify individual studies into 7 different methodology categories described in the database: Probabilistic, Semi-empirical, Model Synthesis, Model Hybrid, Literature Synthesis, Expert Judgement, and Other. Not included on this decision tree are projections for the IPCC category, classified as projections produced from IPCC reports.

Table S2 | Categorization of IPCC emission scenarios for the SLR database

| Emissions Scenario Category | Scenario | Approximate 2100 Temperature (relative to Pre-industrial) |
|---|---|---|
| High | IPCC BAU | 4.3 °C[†] |
|  | IS92e | 3.8 °C[‡] |
|  | SRES A2 | 4.2 °C[§] |
|  | SRES A1FI | 5.0 °C[§] |
|  | RCP8.5 | 4.9 °C[§] |
| Mid | IPCC B | 3.0 °C[†] |
|  | IPCC C | 2.4 °C[†] |
|  | IS92f | 3.5 °C[‡] |
|  | IS92a | 3.0 °C[‡] |
|  | IS92b | 2.9 °C[‡] |
|  | SRES B1 | 2.5 °C[§] |
|  | SRES A1T | 3.0 °C[§] |
|  | SRES B2 | 3.0 °C[§] |
|  | SRES A1B | 3.5 °C[§] |
|  | RCP4.5 | 2.4 °C[§] |
|  | RCP6.0 | 3.0 °C[§] |
| Low | IPCC D | 2.0 °C[†] |
|  | IS92d | 2.2 °C[‡] |
|  | IS92c | 1.9 °C[‡] |
|  | RCP2.6 | 1.5 °C[§] |

[†] Based on values from IPCC FAR (Warrick & Oerlemans, 1990)
[‡] Based on values from IPCC SAR (Warrick et al., 1996), with values adjusted to be relative to pre-industrial by adding 0.61 °C (Hartmann et al., 2013)
[§] Based on median values presented in Table 2 of Rogelj et al. (2012).

## Table S3 | Projected Ranges of SLR from IPCC Reports

| IPCC Report | 2100 SLR Range | Source[*] | End Years | Definition[†] |
|---|---|---|---|---|
| FAR | 0.34 - 0.66 m | FAR Figures 9.6 and 9.7 | 2100 | "Best estimate" range across scenarios |
| FAR | 0.31 - 1.10 m | FAR Figure 9.6 | 2100 | Range for the Policy Scenario Business-as-Usual |
| FAR | 0.16 - 1.10 m[‡] | FAR Figures 9.6 and 9.7 | 2100 | Extreme range of all 4 scenarios |
| SAR | 0.38 - 0.55 m | SAR Summary for Chapter 7 | 2100 | Range of Emission Scenarios IS92a-f using "best estimate" model parameters |
| SAR | 0.20 - 0.86 m | SAR Summary for Chapter 7 | 2100 | Uncertainty range for scenario IS92a |
| SAR | 0.13 - 0.94 m[‡] | SAR Summary for Chapter 7 | 2100 | Extreme range of projections, taking into account both emission scenarios and model uncertainties |
| TAR | 0.09 - 0.88 m[‡] | TAR Executive Summary from Chapter 11 | 2100 | Range of all AOGCMs and SRES scenarios |
| TAR | 0.11 - 0.77 m | TAR Executive Summary from Chapter 11 | 2100 | Range of AOGCMs following the IS92a scenario |
| AR4 | 0.18 - 0.59 m[‡] | AR4 Executive Summary from Chapter 10 | 2090-2099 | Span of the 5-95% range across various SRES scenarios[§] |
| AR5 | 0.26 - 0.82 m[‡] | AR5 Executive Summary from Chapter 13 | 2081-2100 | 'likely' (17th – 83rd percentile) sea-level rise, based on process-based models for all scenarios[**] |
| AR5 | 0.52 - 0.98 m | AR5 Executive Summary from Chapter 13 | 2100 | 'likely' range (17th – 83rd percentile) from process-based models for RCP8.5 |

---

[*] Source from within the IPCC report for the range given
[†] Definition of range from the IPCC report
[‡] Ranges used in discussion of the SLR database
[§] As noted in AR4, these values do not include dynamic ice sheet contributions.
[**] Note that the AR5 report indicates that there is a possibility for an additional contribution to these values of up to several tenths of a meter in the event that the collapse of the marine-based sectors of the Antarctic ice sheet is initiated.

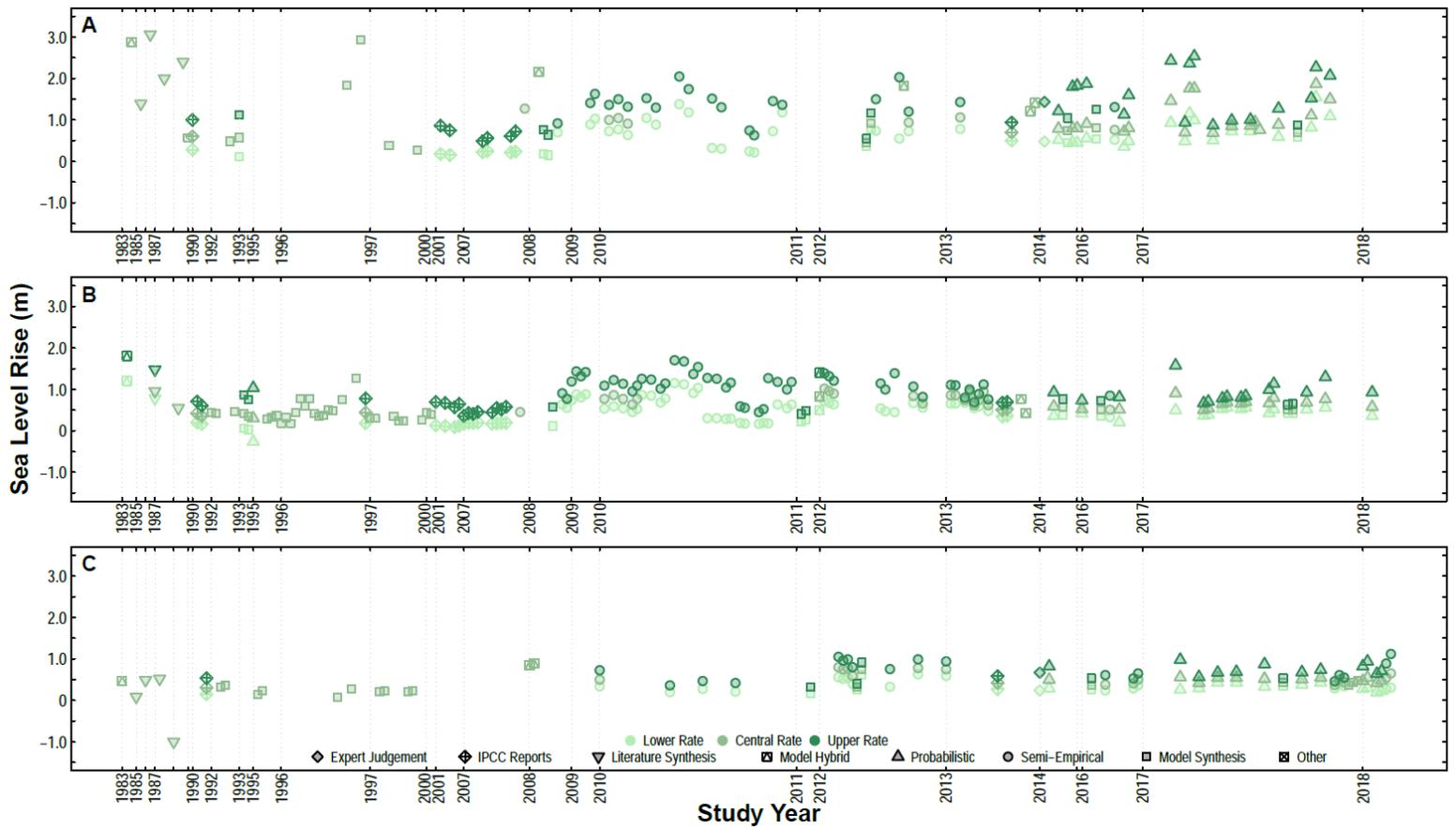

**Figure S2:** Evolution of lower, central, and upper SLR projections from 1983 – 2018. Results are shown for (a) high emissions scenarios, (b) middle emissions scenarios, and (c) low emissions scenarios. Note that time steps are non-uniform, in order to clearly show all projections.

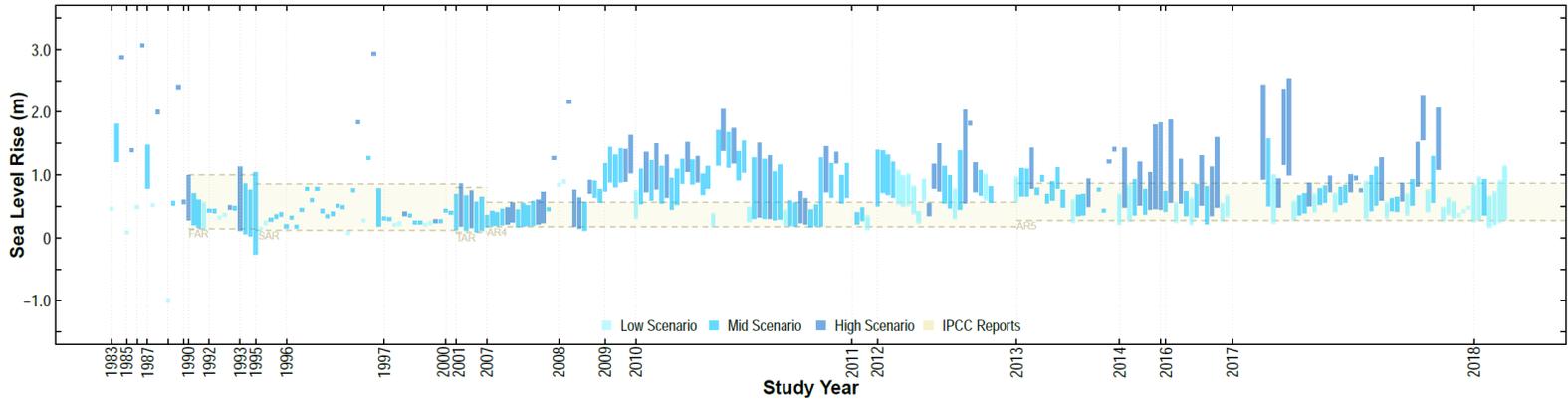

**Figure S3:** Evolution of the ranges of SLR projections throughout time. Length of bars represents the range of each projection made for low emissions scenarios, middle emissions scenarios, and high emissions scenarios. Where possible, bars show the $5^{th}$ – $95^{th}$ percentile range of individual projections from low, middle, and high emissions scenarios. Ranges for IPCC reports (yellow) are as shown in Table S3: the extreme range of projections for IPCC FAR and SAR, the range of all AOGCMs and SRES scenarios for TAR, the 5-95% range across SRES scenarios for AR4 (which do not include dynamic ice sheet response), and the 'likely' ($17^{th}$ – $83^{rd}$ percentile) range from process-based models for AR5 (potential rise above this range as specified in AR5 is not included in the shaded region). Note that time steps are non-uniform, in order to clearly show all projections, and projections have been normalized using Eq. [1] as specified in Section 2.1.

**Table S4 | Median, Likely, and 5th – 95th percentiles of Global Mean Sea Level for Studies Shown in Fig. 5[*]**

| Study | RCP | 2050 5th - 95th Percentile Range (m/50 yrs) | 2050 17th - 83rd Percentile Range (m/50 yrs) | 2050 50th Percentile (m/50 yrs) | 2100 5th - 95th Percentile Range (m/century) | 2100 17th - 83rd Percentile Range (m/century) | 2100 50th Percentile (m/century) |
|---|---|---|---|---|---|---|---|
| Jevrejeva et al., 2012 | RCP3PD | 0.20 - 0.38 | -- | 0.27 | 0.33 - 0.75 | -- | 0.52 |
|  | RCP4.5 | 0.21 - 0.41 | -- | 0.29 | 0.47 - 1.00 | -- | 0.67 |
|  | RCP8.5 | 0.24 - 0.46 | -- | 0.33 | 0.74 - 1.50 | -- | 1.00 |
| Schaeffer et al., 2012 | RCP3PD | -- | -- | -- | 0.52 - 0.96 | -- | 0.75 |
|  | RCP4.5 | -- | -- | -- | 0.64 - 1.21 | -- | 0.90 |
|  | RCP8.5 | -- | -- | -- | 0.72 - 1.39 | -- | 1.02 |
| Perette et al., 2013 | RCP3PD | -- | 0.23 - 0.32 | 0.28 | -- | 0.59 - 0.94 | 0.75 |
|  | RCP4.5 | -- | 0.23 - 0.32 | 0.28 | -- | 0.66 - 1.11 | 0.86 |
|  | RCP8.5 | -- | 0.23 - 0.34 | 0.28 | -- | 0.78 - 1.43 | 1.06 |
| Slangen et al., 2014 | RCP4.5 | -- | -- | -- | -- | 0.37 - 0.77 | 0.57 |
|  | RCP8.5 | -- | -- | -- | -- | 0.45 - 1.04 | 0.75 |
| Kopp et al., 2014 | RCP2.6 | 0.18 - 0.33 | 0.21 - 0.29 | 0.25 | 0.29 - 0.82 | 0.37 - 0.65 | 0.50 |
|  | RCP4.5 | 0.18 - 0.35 | 0.21 - 0.31 | 0.26 | 0.36 - 0.93 | 0.45 - 0.77 | 0.59 |
|  | RCP8.5 | 0.21 - 0.38 | 0.24 - 0.34 | 0.29 | 0.52 - 1.21 | 0.62 - 1.00 | 0.79 |
| Jevrejeva et al., 2014 | RCP8.5 | -- | -- | -- | 0.46 – 1.80 | -- | 0.80 |
| Grinsted et al., 2015 | RCP8.5 | -- | -- | -- | 0.45 - 1.83 | 0.58 - 1.2 | 0.80 |
| Jackson and Jevrejeva, 2016 | RCP4.5 | -- | -- | -- | 0.21 - 0.81 | 0.34 - 0.69 | 0.52 |
|  | RCP8.5 | -- | -- | -- | 0.35 - 1.13 | 0.52 - 0.94 | 0.72 |
|  | "High End" | 0.17 - 0.44 | 0.20 - 0.34 | 0.27 | 0.49 - 1.60 | 0.60 - 1.16 | 0.80 |
| Kopp et al., 2016 | RCP2.6 | -- | -- | -- | 0.24 - 0.61 | 0.28 - 0.51 | 0.38 |
|  | RCP4.5 | -- | -- | -- | 0.33 - 0.85 | 0.39 - 0.69 | 0.51 |
|  | RCP8.5 | -- | -- | -- | 0.52 - 1.31 | 0.59 - 1.05 | 0.76 |
| Mengel et al., 2016 | RCP2.6 | 0.12 - 0.21 | -- | 0.17 | 0.27 - 0.53 | -- | 0.38 |
|  | RCP4.5 | 0.13 - 0.22 | -- | 0.17 | 0.35 - 0.74 | -- | 0.51 |
|  | RCP8.5 | 0.14 - 0.26 | -- | 0.19 | 0.55 - 1.26 | -- | 0.81 |
| Kopp et al., 2017 | RCP2.6 | 0.12 - 0.41 | 0.16 - 0.33 | 0.23 | 0.26 - 0.98 | 0.37 - 0.78 | 0.56 |
|  | RCP4.5 | 0.14 - 0.43 | 0.18 - 0.36 | 0.26 | 0.50 - 1.58 | 0.66 - 1.25 | 0.91 |
|  | RCP8.5 | 0.17 - 0.48 | 0.22 - 0.40 | 0.31 | 0.93 - 2.43 | 1.09 - 2.09 | 1.46 |
| Nauels et al., 2017a | RCP2.6 | -- | 0.17 - 0.27 | 0.22 | -- | 0.34 - 0.54 | 0.43 |
|  | RCP4.5 | -- | 0.19 - 0.28 | 0.23 | -- | 0.43 - 0.63 | 0.52 |
|  | RCP8.5 | -- | 0.20 - 0.30 | 0.25 | -- | 0.58 - 0.87 | 0.71 |
| Nauels et al., 2017b | RCP2.6 | -- | 0.14 - 0.29 | 0.20 | -- | 0.33 - 0.71 | 0.49 |
|  | RCP4.5 | -- | -- | -- | -- | 0.43 - 0.99 | 0.67 |
|  | RCP8.5 | -- | 0.18 - 0.33 | 0.25 | -- | 0.59 - 1.27 | 0.88 |
| Bakker et al., 2017 | RCP2.6 | 0.17 - 0.29 | -- | 0.18 | 0.38 - 0.68 | -- | 0.51 |
|  | RCP4.5 | 0.19 - 0.31 | -- | 0.21 | 0.52 - 0.93 | -- | 0.68 |
|  | RCP8.5 | 0.21 - 0.34 | -- | 0.23 | 0.81 - 1.52 | -- | 1.11 |
| Wong et al., 2017 | RCP2.6 | 0.20 - 0.33 | -- | 0.26 | 0.43 - 0.74 | -- | 0.55 |
|  | RCP4.5 | 0.22 - 0.35 | -- | 0.28 | 0.56 - 1.30 | -- | 0.77 |
|  | RCP8.5 | 0.25 - 0.40 | -- | 0.30 | 1.09 - 2.07 | -- | 1.50 |

---

[*]Note: Projections plotted in Fig. 5 include additional quantiles where available

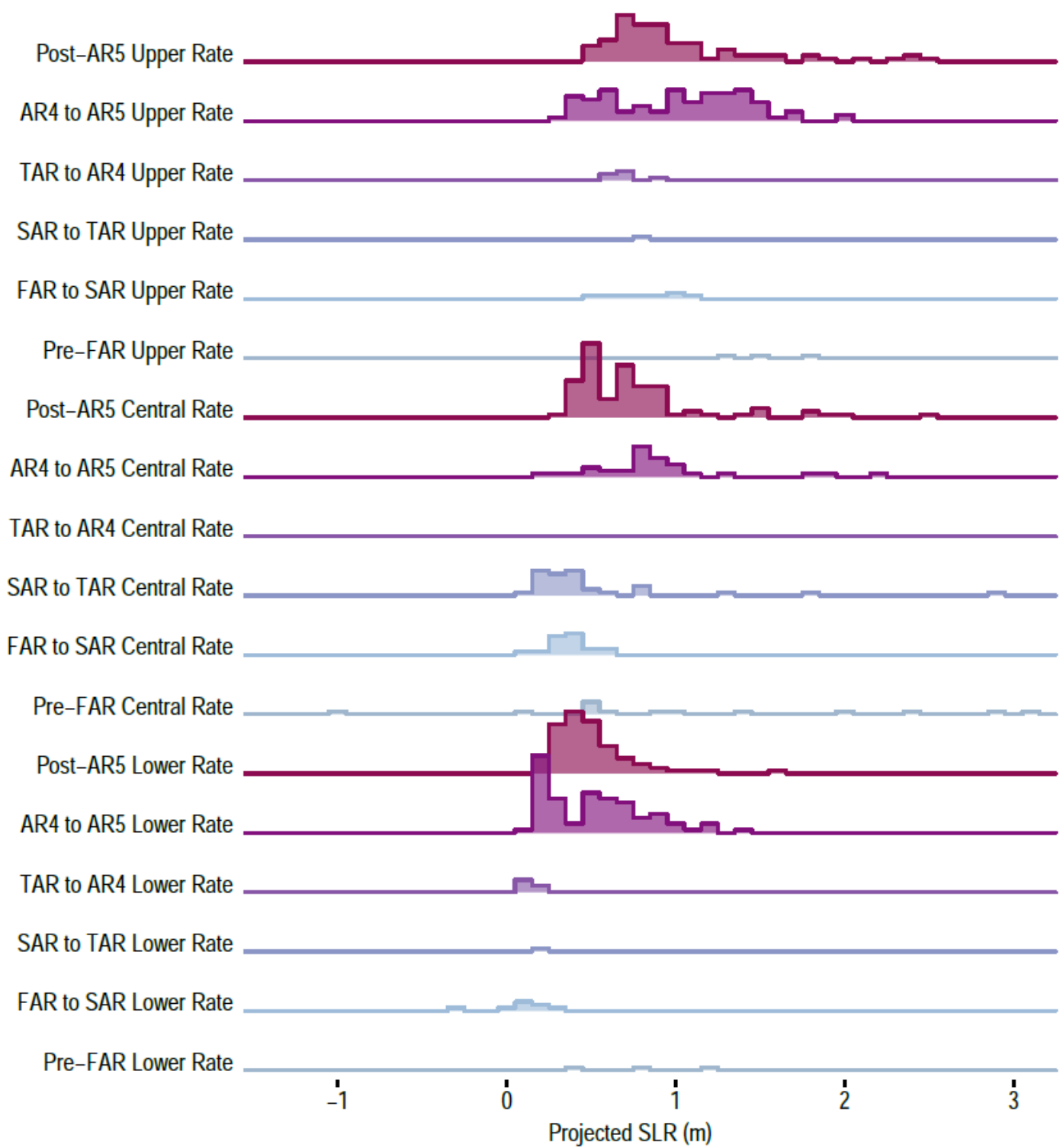

**Figure S4:** Density time series of lower, central, and upper SLR projections. Results are shown for projections made in the time prior to FAR, in the time from FAR to SAR, from SAR to TAR, from TAR to AR4, from AR4 to AR5, and since AR5. Where possible, the 5$^{th}$, 50$^{th}$, and 95$^{th}$ percentile estimates from the original studies are used as lower, central, and upper estimates for each projection included in the time series (see Table S1 and Section 2 for further information about definitions of lower, central, and upper rates).